\documentclass[11pt]{article}

\usepackage{epsfig,graphicx}
\usepackage{amsmath,amsfonts,amssymb}
\usepackage{cite}
\usepackage[usenames,dvipsnames]{color}

\newcommand{\unit}{\leavevmode\hbox{\small1\kern-3.6pt\normalsize1}}

\def \GeV{{\mathrm{GeV}}}

\def\tanb{\tan\beta}

\def\vevs{v_s}
\def\al{{A_\lambda}}
\def\ak{{A_\kappa}}

\def\ln{{\lambda_N}}
\def\aln{{A_{\lambda_N}}}
\def\mn{{m_{\tilde{N}}}}
\def\ayn{A_{y_N}}
\def\yn{y_N}

\def\neut{\widetilde \chi^0}

\def\neut{{\tilde\chi}^0}

\def\neutmass{{m_{{\tilde\chi}_{1}^0}}}

\def\snr{{\tilde N}}

\def\snmassr{{m_{\tilde N_1}}}
\def\snmassrsq{{m^2_{\tilde N_1}}}

\def\rhn{{N}}
\def\rhnmass{{M_{N}}}

\newcommand{\higgs}[1]{H_{#1}^0}
\newcommand{\phiggs}[1]{A_{#1}^0}

\def\hsm{h_{SM}^0}

\def\higgsi{{H_i^0}}

\def\higgsl{{H_1^0}}

\def\hmassj{m_{H_j^0}}

\def\hmassl{m_{H_1^0}}
\def\hmassm{m_{H_2^0}}

\def\hcompid{S_{H_i^0}^1}
\def\hcompiu{S_{H_i^0}^2}
\def\hcompis{S_{H_i^0}^3}

\def\phiggsl{{A_1^0}}

\def\phmassl{{m_{A_1^0}}}
\def\phmassh{{m_{A_2^0}}}

\def\chiggspm{{H^\pm}}

\def\chsnrsnr{C_{\higgsi\snr_1\snr_1}}

\newcommand{\sigsi}{\sigma^{SI}_{\snr_1 p}}

\newcommand{\bsg}{b\to s\gamma}
\newcommand{\bmumu}{B_S\to\mu^+\mu^-}
\newcommand{\asusy}{a^{\rm SUSY}_\mu}

\def\lsim{\raise0.3ex\hbox{$\;<$\kern-0.75em\raise-1.1ex\hbox{$\sim\;$}}}
\def\gsim{\raise0.3ex\hbox{$\;>$\kern-0.75em\raise-1.1ex\hbox{$\sim\;$}}}

\allowdisplaybreaks

\begin{document}

\thispagestyle{empty}
\begin{flushright}
  IFT-UAM/CSIC-14-031\\
  FTUAM-14-14\\

  \vspace*{2.mm}{April 9, 2014}
\end{flushright}

\begin{center}
  {\Large \textbf{Low-mass right-handed sneutrino dark matter: \\
  SuperCDMS and LUX constraints and the Galactic Centre gamma-ray excess  } }  
    
  \vspace{0.5cm}
  D.G.~Cerde\~no,
  M.~Peir\'o$^{a}$, and
  S.~Robles
  \\[0.2cm] 
    
  {\footnotesize{ 
      Instituto de F\'{\i}sica Te\'{o}rica
      UAM/CSIC, Universidad Aut\'{o}noma de Madrid,  28049
      Madrid, Spain\\[0pt] 
      Departamento de F\'{\i}sica Te\'{o}rica,
      Universidad Aut\'{o}noma de Madrid, 28049
      Madrid, Spain\\[0pt] 
        }
    }
  
\vspace*{0.7cm}
\begin{abstract}
\footnotesize{
Recent results from direct and indirect searches for dark matter (DM) have motivated the study of particle physics models that can provide weakly interacting massive particles (WIMPs) in the mass range $1-50$~GeV. 
Viable candidates for light WIMP DM must fulfil stringent constraints. 
On the one hand, the observation at the LHC of a Higgs boson with Standard Model properties set an upper bound on the coupling of light DM particles to the Higgs, thereby making it difficult to reproduce the correct relic abundance.
On the other hand, the recent results from direct searches in the CDMSlite, SuperCDMS and LUX experiments have set upper
constraints on the DM scattering cross section.
In this paper, we investigate the viability of light right-handed sneutrino DM in the Next-to-Minimal Supersymmetric Model (NMSSM) in the light of these constraints. 
To this aim, we have carried out a scan in the NMSSM parameter space, imposing experimental bounds on the Higgs sector
and low-energy observables, such as the muon anomalous magnetic moment and branching ratios of rare decays. 
We demonstrate that the enlarged Higgs sector of the NMSSM, together with the flexibility provided by the RH sneutrino parameters, make it possible to obtain viable RH sneutrino DM with a mass as light as $2$~GeV. 
We have also considered the upper bounds on the annihilation cross section from Fermi LAT data on dwarf spheroidal galaxies, and extracted specific examples with mass in the range $8-50$ GeV that could account for the apparent low-energy excess in the gamma-ray emission at the Galactic Centre.
Then, we have computed the theoretical predictions for the elastic scattering cross-section of RH sneutrinos. Finally, after imposing the recent bounds from SuperCDMS and LUX, we have found 
a wide area of the parameter space that could be probed by future low-threshold direct detection experiments.  
}
\end{abstract}
\end{center}

\let\oldthefootnote\thefootnote
\renewcommand{\thefootnote}{\alph{footnote}}
\footnotetext[1]{MultiDark Scholar}
\let\thefootnote\oldthefootnote
\setcounter{footnote}{0}

\newpage
	

\section{Introduction} 
\label{sec:intro}

Weakly-interacting massive particles (WIMPs) are appealing candidates to account for the dark matter (DM) of the Universe, since they can be thermally produced in a quantity that agrees well with current observations. 
WIMPs can be searched for through their scattering off nuclei in the targets of underground direct detection experiments. Other experiments are looking for the annihilation products of WIMPs in the galactic halo as 
indirect evidence of the DM paradigm. Finally, the Large Hadron Collider (LHC) is probing  models for new physics at the TeV scale, thereby 
putting many WIMP scenarios for DM to the test.

Various direct detection experiments have reported signals that could be 
interpreted in terms of light WIMPs with a mass of order 5-20~GeV. In particular, the DAMA collaboration observed an annual modulation in the detection rate \cite{Bernabei:2003za} that has been later confirmed with an extended version of the experiment, DAMA/LIBRA \cite{Bernabei:2008yi}.
Later, the CoGeNT \cite{Aalseth:2010vx,Aalseth:2012if} and CRESST \cite{Angloher:2011uu} collaborations also reported irreducible excesses over their known backgrounds.
In the case of CoGeNT, the signal has also been observed to display an annual modulation with a significance of order $2\,\sigma$  \cite{Aalseth:2011wp,Aalseth:2014nda}, 
and the latest data would be compatible with a WIMP mass of order $12$~GeV and a WIMP-nucleon elastic scattering cross-section of order $\sigma^{SI}\approx2\times10^{-6}$ pb \cite{Aalseth:2014jpa} (somewhat lower than previously observed \cite{Aalseth:2012if}).
Adding excitement to the field, a recent analysis of the data taken with the silicon detectors of the CDMS II experiment \cite{Agnese:2013cvt,Agnese:2013rvf} resulted in three candidate events. Despite not being statistically significant to make any discovery claim, the reconstruction of WIMP parameters corresponding to the CDMS II signal could also point towards light WIMPs.

However, these results are in conflict with the non-observation of DM signals in other detectors,
such as SIMPLE \cite{Felizardo:2011uw}, KIMS \cite{Kim:2012rz}, a combination of CDMS and EDELWEISS data \cite{Ahmed:2011gh}, XENON10 \cite{Angle:2011th}, and XENON100 \cite{Aprile:2011hi,Aprile:2012nq}, from which upper constraints on $\sigma^{SI}$ can be derived.
The most stringent bounds have been obtained with LUX \cite{Akerib:2013tjd}, CDMSlite \cite{Agnese:2013jaa} and SuperCDMS \cite{Agnese:2014aze} experiments.
LUX excludes $\sigma^{SI}\gsim10^{-9}$~pb for WIMPs with a mass of order 30~GeV. On the other hand, CDMSlite and SuperCDMS dominate for low-mass WIMPs, setting  
$\sigma^{SI}\gsim10^{-5}$~pb for WIMPs with $m_{DM}\approx6$~GeV.
The situation regarding light WIMPs might be clarified in the near future, since several direct detection experiments will be able to further probe the low-mass window more precisely either using light targets, but mostly through a reduction in the energy threshold. Such could be the case of the SuperCDMS proposal for a detector in SNOLAB, which could probe light WIMPs with a scattering cross-section as small as approximately $10^{-8}$~pb for $m_{DM}\lesssim8$~GeV. 
In this sense, it is necessary to understand which theoretical models can be tested.

Further motivation for very light WIMPs has been found in data from indirect dark matter searches. 
In particular, it has been argued that Fermi LAT observations of the gamma ray flux from the Galactic Centre (GC) show a small excess at low energies that could 
be explained in terms of DM particles \cite{Vitale:2009hr,Morselli:2010ty,Hooper:2010mq,
Hooper:2011ti,Abazajian:2012pn,Hooper:2013nhl,Gordon:2013vta,Abazajian:2014fta} with a mass in the range $10-60$~GeV, depending on the specific annihilation channel, and an annihilation cross-section
which is remarkably close to that expected for a thermal WIMP.
For instance, the observed spectrum is well fit by $31- 40$ GeV DM particles annihilating to $b\bar b$ \cite{Daylan:2014rsa}.
Nevertheless, these studies involve the subtraction of an astrophysical background on which there might be significant uncertainties. 
In fact, indirect detection experiments have provided data that constrain low mass WIMPs.
The results of Fermi LAT searches for gamma rays in dwarf spheroidal (dSph) galaxies and the centre of the Galaxy can be interpreted as a lower 
bound on ``thermal'' DM candidates (i.e., those for which the annihilation cross-section in the early Universe is $\langle\sigma v\rangle\approx3\times 10^{-26}$ cm$^3$s$^{-1}$) 
\cite{GeringerSameth:2011iw,Ackermann:2011wa,Hooper:2011ti,Hooper:2012sr,Gomez-Vargas:2013bea,Ackermann:2013yva} that constrain especially light WIMPs. 
Similar bounds have also been obtained from studies of the diffuse gamma ray emission \cite{Abdo:2010dk,Ackermann:2012rg,Ackermann:2012qk,fortheFermiLAT:2013naa,Tavakoli:2013zva} 
and from an analysis of the Andromeda galaxy \cite{Li:2013qya}.

Constraints on light WIMPs can also be derived from the results of the first years of operation of the Large Hadron Collider (LHC). 
The observation of a Higgs boson with a mass of approximately $126$~GeV \cite{Higgs,CMS:2012gu,ATLAS:2013mma,CMS:yva} and couplings very similar to those predicted by the SM entails an upper limit on the invisible branching ratio of the SM Higgs, BR$(\hsm\to inv)\lesssim 20\%-30\%$ \cite{ATLAS:2013pma,
Espinosa:2012vu,Belanger:2013kya,Falkowski:2013dza,Giardino:2013bma,Ellis:2013lra,Djouadi:2013qya,Belanger:2013xza,Bechtle:2014ewa}, which applies to the possible decay into light DM particles, limiting their coupling to the SM Higgs boson. 
Due to the correlation between diagrams that contribute to the DM annihilation (and therefore relic density) and direct detection, 
it has been shown in several examples that if the SM Higgs is the only mediator between the DM and the SM particles (e.g., a Higgs portal) 
then the current constraint on BR$(\hsm\to \chi\chi)$ rules out very light WIMPs (which would otherwise be consistent with CDMS II Silicon and CoGeNT data). 
This correlation can be broken in models with an extended Higgs sector (e.g., non-minimal Supersymmetric models) in which the DM couples to a non-SM-like Higgs. 
This is the case of the Next-to-minimal Supersymmetric Standard Model (NMSSM).

WIMP DM can be easily accommodated in particle models for physics beyond the Standard Model (SM). 
A paradigmatic example is Supersymmetry (SUSY), a framework in which the lightest supersymmetric particle (LSP) can be stable if 
R-parity is unbroken. The lightest neutralino, ${\tilde\chi_1^0}$, is generally considered  as the natural DM candidate.
It has been shown that in specific corners of the parameter space of the Minimal Supersymmetric Standard Model (MSSM) the experimental bounds can be circumvented in order to obtain solutions with  $m_{\tilde\chi_1^0}\gsim 10-15$~GeV 
\cite{Pierce:2013rda,Hagiwara:2013qya,Belanger:2013pna,Boehm:2013qva}. 
In the Next-to-Minimal Supersymmetric Standard Model (NMSSM), the extended Higgs sector makes it possible to lower this bound to $m_{\tilde\chi_1^0}\sim5$~GeV \cite{AlbornozVasquez:2011js,Kozaczuk:2013spa,Cao:2013mqa}. 
However, in both scenarios the regions of the parameter space in which these 
can appear are very fine-tuned, mostly in order to obtain the correct relic abundance.

In this article, we consider an extension of the NMSSM with RH neutrino and sneutrino states \cite{Cerdeno:2008ep,Cerdeno:2009dv}. 
This construction features two singlet superfields, as in 
Refs.\,\cite{ko99,pilaftsis}. A singlet superfield, $S$, is the usual NMSSM scalar Higgs which 
addresses the $\mu$ problem  \cite{Kim:1983dt}
and provides extra Higgs and neutralino states, while an additional singlet
superfield, $N$, accounts for RH neutrino and sneutrino
states. 
In this scenario, the RH sneutrino can be a viable WIMP dark matter candidate for a wide range of masses \cite{Cerdeno:2008ep,Cerdeno:2009dv} provided that it 
is the lightest supersymmetric particle (LSP). 
In particular, in Ref.\,\cite{Cerdeno:2011qv}, it was shown that light RH sneutrinos, with masses below $10$~GeV could be obtained.

In the light of the experimental results enumerated above, we study the RH sneutrino phenomenology in the NMSSM as a light
DM candidate. 
We demonstrate that the enlarged Higgs sector of the NMSSM, together with the flexibility provided by the RH sneutrino parameters, make it possible to obtain 
viable RH sneutrino DM with a mass as low as 2 GeV. We obtain specific examples for which a RH sneutrino with a mass in the range $8-50$~GeV can account 
for the apparent low-energy excess in the gamma-ray emission at the Galactic Centre. 
Finally, we evaluate the impact of the recent results from direct DM searches in the LUX and SuperCDMS experiments. We show that a wide area of the 
parameter space survives which could be probed by future low-threshold experiments.

This paper is organised as follows. In Section \ref{sec:model}, we review the basic features of the NMSSM with RH 
sneutrinos and briefly describe some aspects about its DM phenomenology. In Section \ref{sec:verylight}, we investigate the 
viability of the RH sneutrino as a low-mass WIMP. With this purpose, we perform a scan over some parameters of the model 
and identify the dominant annihilation channels, 
taking into account the most recent experimental constraints from the Higgs sector, low energy observables and DM abundance. We also compute the spin-independent scattering 
cross-section of the RH sneutrino-nucleon. Afterwards, we analyse the resulting configurations of the parameter space and 
compare them with the exclusion limits set by direct and indirect detection experiments.
Finally, in Section \ref{sec:conclusions} we present our conclusions.

\section{Right-handed sneutrino in the NMSSM}
\label{sec:model}

The model, described in Refs.~\cite{Cerdeno:2008ep,Cerdeno:2009dv}, is an extended version of the NMSSM, in which a right-handed neutrino superfield $N$, singlet under the SM gauge group, is added in order to account for RH neutrino and sneutrino states. The superpotential of this construction is given by
\begin{eqnarray}
  W &=& W_{\rm NMSSM} + \lambda_N S N N + y_N L \cdot H_2 N, 
    \label{eq:superpotential}
\end{eqnarray}
where flavour indices are omitted and the dot denotes the antisymmetric $SU(2)_L$ product. $ W_{\rm NMSSM}$ is the NMSSM superpotential, $\ln$ is a new dimensionless coupling, 
$y_N$ is the neutrino Yukawa coupling, and $H_{1,2}$ are the down and up type doublet Higgs components, respectively.
As in the NMSSM, a global $Z_3$ symmetry is imposed so that there are no supersymmetric mass terms in the superpotential.
Since we assume $R$-parity conservation in order to guarantee the stability of the LSP, the terms $NNN$ and $SSN$ are forbidden. Furthermore, we do not consider CP violation in the Higgs sector\footnote{Spontaneous CP-violation in this model has been studied in Ref.~\cite{Huitu:2012rd} and later applied to the emission of a monochromatic photon line~\cite{Chatterjee:2014bva}.}.

The Lagrangian, with the corresponding soft SUSY-breaking terms, reads
\begin{equation}
	-{\cal L} = -{\cal L}_{\rm NMSSM}
	+ \mn^2 |\tilde{N}|^2  + \left( 
	\ln\aln S \tilde{N}^2 + \yn\ayn \tilde{L} H_2 \tilde{N}+ {\rm H.c.}  \right) ,
	\label{lagrangian_couplings}
\end{equation}
where ${\cal L}_{\rm NMSSM}$ is the NMSSM Lagrangian, to which we add a soft mass term for the RH sneutrino, $\mn$, and two new trilinear soft terms $\aln$ and $A_{\yn}$.

After radiative electroweak symmetry-breaking the Higgs fields get non-vanishing vacuum expectation values (VEVs).
 The physical CP-even and CP-odd Higgs eigenstates can be expressed as a linear superposition of the $H_d$, $H_u$ and $S$ fields. For the CP-even Higgs we will use the following decomposition,
\begin{equation}
\higgsi = \hcompid H_d + \hcompiu H_u + \hcompis S\, . 
\end{equation}

The VEV of
the singlet Higgs, $\vevs$, induces an effective $\mu$ parameter, $\mu=\lambda\vevs$, and a Majorana mass term for the RH neutrinos, $\rhnmass=2\ln\vevs$, both of which are of the order of the electroweak scale \cite{Cerdeno:2008ep}. 
In order to reproduce the light left-handed (LH) neutrino masses, the see-saw mechanism relation implies that $\yn$ has to be similar to the electron Yukawa, $\yn\sim10^{-6}$, 
typical value of a low-scale see-saw mechanism.
This leads to a very small left-right mixing in both the neutrino and sneutrino sectors and consequently the mass eigenstates can be identified with the LH and RH components.
The smallness of $\yn$ has interesting implications for collider physics, such as the occurrence of displaced vertices (due to the late decay of RH neutrinos) or charged tracks (from long-lived staus) \cite{Cerdeno:2013oya}, 
but it is otherwise not important for DM phenomenology.

Regarding the sneutrino sector, the lighter RH sneutrino mass, $\snmassr$, can be expressed in terms
of the rest of the parameters as follows \cite{Cerdeno:2011qv},
\begin{equation}
	\snmassrsq= m_{\tilde{N}}^2 +|2\ln \vevs|^2  + |y_N v_2|^2 
	\pm 2 \ln \left( A_{\lambda_N} \vevs+
	(\kappa \vevs^2-\lambda v_1 v_2 )^{\dagger} \right) ,
	\label{eq:snmass}
\end{equation}
where the sign in front of $2\ln$ is chosen opposite to the sign of $\left(A_{\lambda_N} \vevs+(\kappa \vevs^2-\lambda v_1 v_2 )^{\dagger} \right)$.
It was shown in Refs.\cite{Cerdeno:2008ep,Cerdeno:2009dv} that the RH sneutrino can be a viable candidate for WIMP DM, reproducing the correct relic abundance for a wide range of masses, including cases in which the RH sneutrino is very light \cite{Cerdeno:2011qv}. 
The flexibility of this construction stems from the fact that the new free parameters 
$(\ln,\mn,\,\aln)$ can be chosen to fix the RH sneutrino mass and its coupling to the singlet Higgs boson without affecting the rest of the NMSSM spectrum. 
Without loss of generality, we can choose the physical mass of the RH sneutrino, $\snmassr$, as a free parameter 
(and determine the soft mass accordingly). 
This is the approach that we have followed in this paper, hence our sneutrino parameter space is defined by $(\ln,\snmassr,\,\aln)$.

The RH sneutrino couplings to Higgses and
neutralinos arise from the $SNN$ term in the superpotential and Lagrangian.
The strength of the interaction is therefore dependent on the value of $\ln$ and 
$\aln$. 
For suitable values of $\ln$ and $\aln$, this interaction is of the order of the electroweak scale, 
thereby making the RH sneutrino a WIMP candidate. The RH sneutrino annihilation channels include at tree level the following possibilities:
\begin{description}
\item {(i)}
  $W^+\, W^-$, $Z\, Z$, and $f \bar{f}$
  via $s$-channel Higgs exchange;
\item {(ii)}
  $H_i^0\, H_j^0$, 
  via $s$-channel Higgs exchange, $t$-
  and $u$-channel sneutrino exchange, and a scalar quartic coupling;
\item {(iii)}
  $A_a^0\, A_b^0$, and $H_i^+\, H_j^-$,  
  via  $s$-channel Higgs exchange, and a scalar quartic coupling;
\item {(iv)}
  $Z\,A_a^0$ and $W^\pm\,H^\mp$
  via $s$-channel Higgs exchange;
\item {(v)}
  $NN$, 
  via $s$-channel Higgs exchange and via $t$- and $u$-channel
  neutralinos exchange.
\end{description}
The processes suppressed by the neutrino Yukawa (such as
$s$-channel sneutrino annihilation mediated by the $Z$ boson) 
have not been included, since they are negligible.

Other annihilation products at one-loop include gluons and photons,
\begin{description}
\item {(vi)}
  $g g$, via $s$-channel Higgs exchange with a quark loop (mainly top, bottom and charm) and a squark loop;
\item {(vii)}
  $\gamma\gamma$, $Z\gamma$, 
  via a loop of $\chiggspm$
  and $s$-channel Higgs exchange with a loop of quarks, $W^\pm$, $\chiggspm$, charginos 
  and sfermions.
\end{description}

\section{Light RH sneutrino dark matter in the NMSSM}
\label{sec:verylight}

We have carried out a scan in the parameter space of this model in order to look for solutions with light RH sneutrino DM.
The parameters of the model that determine the phenomenology of the RH sneutrino are varied in the ranges detailed in Table~\ref{tab:parameters}. 
Out of convenience, we have split the scan in two regions in $\tanb$, namely $\tan\beta=[4 , 10]$ and $\tan\beta=[10 , 20]$.
The rest of the input parameters have less impact on the RH sneutrino properties, but do affect the theoretical predictions for low-energy observables. 
Hence, they are considered fixed to some suitable values. 
In particular, gaugino soft masses are taken to be $M_1=350~\GeV$, $M_2=700~\GeV$ and $M_3=2100~\GeV$, thus satisfying the Grand Unification relation. Slepton and squark soft masses are equal for the three families, 
$m_{\widetilde{L}}=m_{\widetilde{e}^c}=300~\GeV$, and
$m_{\widetilde{Q}}=m_{\widetilde{u}^c}=m_{\widetilde{d}^c}=1500~\GeV$.
Trilinear soft terms are chosen to be $A_{t}=3700~\GeV$, $A_{b}=2000~\GeV$, $A_{\tau}=-1000~\GeV$.
All these parameters are defined at the EW scale.

We have implemented this construction in {\tt CalcHEP 3.4.3}~\cite{Belyaev:2012qa} model files so that we can calculate 
the RH sneutrino relic abundance with {\tt micrOMEGAs 3.2} \cite{Belanger:2013oya}.
We use {\tt NMSSMTools 4.0.0} \cite{Ellwanger:2004xm,Ellwanger:2005dv,Ellwanger:2006rn} to compute the NMSSM mass spectrum, the masses of Higgs bosons including full two-loop contributions, and the relevant 
low-energy phenomenology observables.
In order to refine our exploration of the parameter space, we have linked those codes with {\tt MultiNest 3.0} \cite{Feroz:2007kg,Feroz:2008xx}. 
To that end, we have built a likelihood function, whose 
parameters are the CDM relic density, $m_{\hsm}$, ${\rm BR}(B_s\to \mu^+\mu^-)$, and ${\rm BR}(b\to s\gamma)$, 
which are taken as gaussian probability distribution functions around the measured values with 
$2\sigma$ deviations.  This likelihood function is 
used to generate MCMC chains and to find regions of the 
parameter space that maximise the likelihood. Using {\tt MultiNest} allows 
us to scan the parameter space of the model more efficiently, since relatively 
few evaluations are needed to converge to regions of maximum likelihood.

\begin{table}[!t]
  \begin{center}
    \begin{tabular}{|ccccc|}
      \hline
     & Parameter & & Range & \\
      \hline
     & $\tanb$ & & $[4 , 10]$, $[10 , 20]$  & \\
     & $\lambda$& & $[0.1, 0.6]$& \\
     & $\kappa$& & $[0.01, 0.1]$ & \\
     & $\al$& & $[500 , 1100]$ & \\
     & $\ak$& & $[-50 , 50]$ & \\
     & $\mu$& & $[110 , 250]$ & \\
     & $\ln$& & $[0.07 , 0.4]$ & \\
     & $\aln$& & $[-1100 , -500]$ & \\
     & $\snmassr$& &$[1 , 50]$ & \\     
      \hline
    \end{tabular}
    \vspace{0.4cm}
    \caption{\small Ranges of variation of the input parameters used in the scan. 
    Masses and trilinear terms are given in GeV units. All the parameters are defined at the EW scale.}
    \label{tab:parameters}
  \end{center}
\end{table}

\subsection{Experimental constraints}
\label{sec:constraints}

Low-energy observables have an important impact in the allowed regions of the NMSSM sector. 
We have implemented the recent measurement of the branching ratio of the $B_s\to \mu^+\mu^-$ process
by the LHCb \cite{Aaij:2013aka} and CMS \cite{Chatrchyan:2013bka} collaborations, which collectively yields
$1.5\times 10^{-9}< {\rm BR}(B_s\to \mu^+\mu^-)< 4.3\times 10^{-9}$ at
95\% CL. 
This experimental constraint is relatively easy to fulfil in our analysis since it depends on the mass of the second pseudoscalar, $\phmassh\gsim 500$~GeV, 
and $\tan\beta$, which is moderately large, $\lsim20$.
Even if the lightest pseudoscalar can be very small in our scan, 
it is an almost pure singlet, thereby not inducing new contributions to BR($\bmumu$).
For the $b\to s\gamma$ decay, we have considered the 2$\sigma$ range
$2.89\times 10^{-4}< {\rm BR}(b\to s\gamma)< 4.21\times
10^{-4}$, which takes into account
theoretical and experimental uncertainties added in quadrature \cite{Ciuchini:1998xy,D'Ambrosio:2002ex,Misiak:2006zs,
Misiak:2006ab,Amhis:2012bh}.
The effect of this constraint on the NMSSM parameter space can be rather severe for small values of $\tan\beta$ \cite{Cerdeno:2007sn,Ellwanger:2009dp}. 
Then, we have set the values of the squark sector (mainly the sbottom mass parameters and $A_b$) in such a way that this bound is satisfied. 
In addition, we have imposed $0.85\times 10^{-4}< {\rm BR}(B^+ \to \tau^+ \nu_\tau)<2.89\times 10^{-4}$ \cite{Lees:2012ju}.

Concerning the muon anomalous magnetic moment, experimental
results using $e^+e^-$ data show a discrepancy with the SM prediction 
\cite{Bennett:2006fi,Jegerlehner:2009ry,Gray:2010fp,Davier:2010nc,Hagiwara:2011af} that can be interpreted as a hint of SUSY. From   $e^+e^-$ data the SUSY contribution is constrained to be $10.1\times 10^{-10}<a_{\mu}^{\rm SUSY}<42.1\times10^{-10}$ at $2\sigma$, although tau data suggest a smaller discrepancy,  
$2.9\times 10^{-10}<a_{\mu}^{\rm SUSY}<36.1\times10^{-10}$ at $2\sigma$  \cite{Davier:2010nc}.
A recent update using the Hidden Local Symmetry (HLS) model leads to the combined $(e^+e^-+\tau)$ result $16.5\times 10^{-10}<a_{\mu}^{\rm SUSY}<48.6\times10^{-10}$ at $2\sigma$  \cite{Benayoun:2012wc}. In our analysis we will quote these three ranges. 
To ensure a sizable contribution to this observable we have fixed the values of slepton soft masses $m_{\widetilde{L}}=m_{\widetilde{e}^c}=300~\GeV$ near their experimental lower limit.  
Also, large $\tan\beta$ regions are favoured by this constraint. 
It is worth noting that the computation of the low-energy observables is performed as in the NMSSM, since there is no significant contribution from the RH sneutrino sector.

Regarding the Higgs sector, we have imposed the presence of a SM-like Higgs,
in the mass range $123-128$~GeV.
In our scan the SM-like Higgs corresponds to the second mass eigenstate, $\higgs{2}$, as a lighter singlet-like state $\higgs{1}$ is needed to couple to RH sneutrinos without violating experimental bounds.
In order to maximise one-loop contributions to the SM-like Higgs mass, we have set large values for the squark soft masses, $m_{\widetilde{Q},\widetilde{u}^c,{d}^c}$, the gluino mass term, $M_3$ and the top trilinear parameter, $A_t$.
For the reduced
signal strength
of the Higgs to di-photon mode, $R_{\gamma\gamma}$, we have used
$0.23\leq R_{\gamma\gamma}\leq 1.31$, the latest CMS results at
2$\sigma$~\cite{CMS:yva}\footnote{For ATLAS the same
limit including all systematics is $0.95\leq R_{\gamma\gamma}\leq 2.55$ \cite{ATLAS:2013oma,ATLAS:2013wla}.}.
The remaining reduced signal strengths are also constrained according
to the CMS results of Ref.~\cite{CMS:yva} (see Refs.~\cite{ATLAS:2013mma,ATLAS:2013wla} for the equivalent ATLAS results).
Notice that these measurements indirectly entail a
strong bound on the invisible and non-standard
decay modes of the SM-like Higgs boson~\cite{ATLAS:2013pma,
Espinosa:2012vu,Belanger:2013kya,Falkowski:2013dza,Giardino:2013bma,Ellis:2013lra,Djouadi:2013qya,Belanger:2013xza}, which
in our case affects the
$\higgs{2} \to \snr_1\snr_1$,  $\higgs{2} \to \higgs{1}\higgs{1}$, 
$\higgs{2} \to \phiggs{1}\phiggs{1}$, $\higgs{2} \to \neut_1\neut_1$ and  $\higgs{2} \to \rhn\rhn$ decay modes.
Using the $2\sigma$ limit derived in Ref.~\cite{Cerdeno:2013oya}, we have considered ${\rm BR}(\hsm\rightarrow {\rm inv})<0.27$, consistent with other recent analyses \cite{Espinosa:2012vu,Belanger:2013kya,Falkowski:2013dza,Giardino:2013bma,Ellis:2013lra,Djouadi:2013qya,Belanger:2013xza}. 

Finally, we have required the lightest sneutrino to be the lightest supersymmetric particle (LSP) and set
an upper bound on its relic abundance, $\Omega_{\snr_1} h^2<0.13$,
consistent with the latest Planck results~\cite{Ade:2013zuv}. 
Besides, we have considered the possibility that RH sneutrinos only contribute to a fraction of the total relic density, and set for concreteness a lower bound on the relic abundance, $0.001<\Omega_{\snr_1} h^2$. 
To deal with these cases, the fractional density, $\xi=\min[1,\Omega_{\snr_1}h^2/0.11]$, will be introduced to account for the reduction in the rates for direct and indirect searches (assuming that the RH sneutrino is present in the DM halo in the same proportion as in the Universe). 
We will investigate the effect of the recent bounds that can be derived from direct and indirect dark matter searches in Sections \ref{sec:direct} and \ref{sec:indirect} respectively.
More specifically, we will apply the LUX \cite{Akerib:2013tjd}, CDMSlite \cite{Agnese:2013jaa}, and SuperCDMS \cite{Agnese:2014aze}
upper bounds on the RH sneutrino scattering cross-section off nucleons. 
With respect to indirect detection, we will take into account the upper constraints on the gamma ray flux from RH sneutrino annihilation cross-section set by Fermi LAT data \cite{GeringerSameth:2011iw,Ackermann:2011wa}.

\subsection{Annihilation channels}
\label{sec:relic}

Reproducing the correct relic abundance is generally very complicated for light DM candidates, since many annihilation channels are not kinematically open. On top of this, the recent constraints on the Higgs sector limit the coupling of the SM Higgs to light DM particles (since they contribute to the Higgs invisible branching ratio), preventing them from having a sufficiently large annihilation cross-section. This is particularly important in models with only the SM Higgs or in which the lightest Higgs is SM-like (such as in the case of the neutralino in the MSSM).

These difficulties are easy to overcome in our model. First, the Higgs sector in the NMSSM is extended and offers the possibility of having a light singlet-like Higgs which can contribute significantly to the DM couplings with SM particles while simultaneously presenting a $126$~GeV Higgs boson with SM couplings \cite{Vasquez:2012hn}.  
In addition, since light scalar and pseudoscalar Higgs bosons are viable, 
new annihilations channels open up, as well as the possibility of having a resonance effect when $2\snmassr\approx\hmassl$.
Finally, the RH sneutrino mass can be selected through the new inputs of the model (it can be considered a free parameter) without affecting the rest of the NMSSM spectrum.

\begin{figure}[!t]
	\begin{center}  
	\epsfig{file=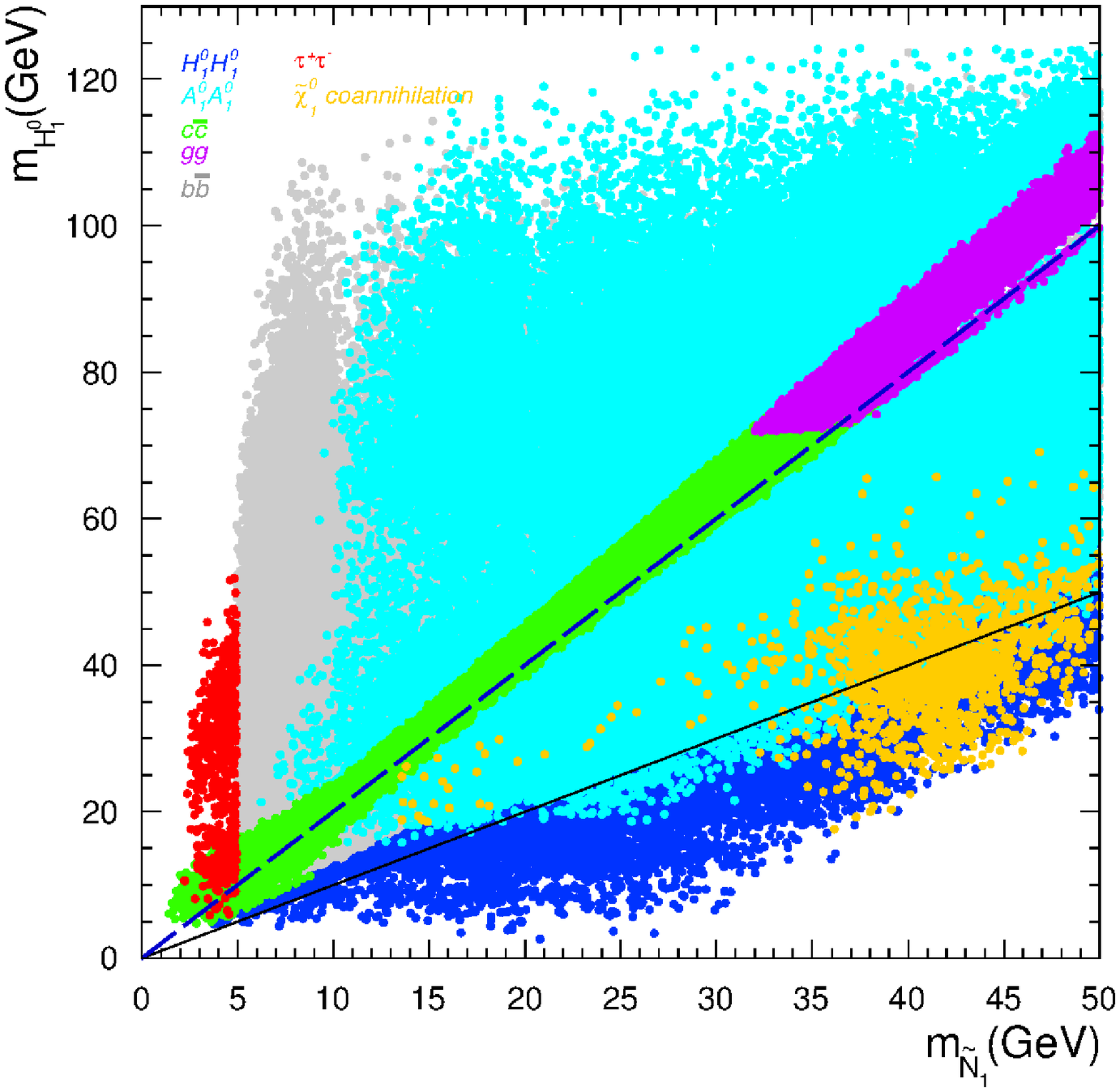,width=7.5cm}
	\epsfig{file=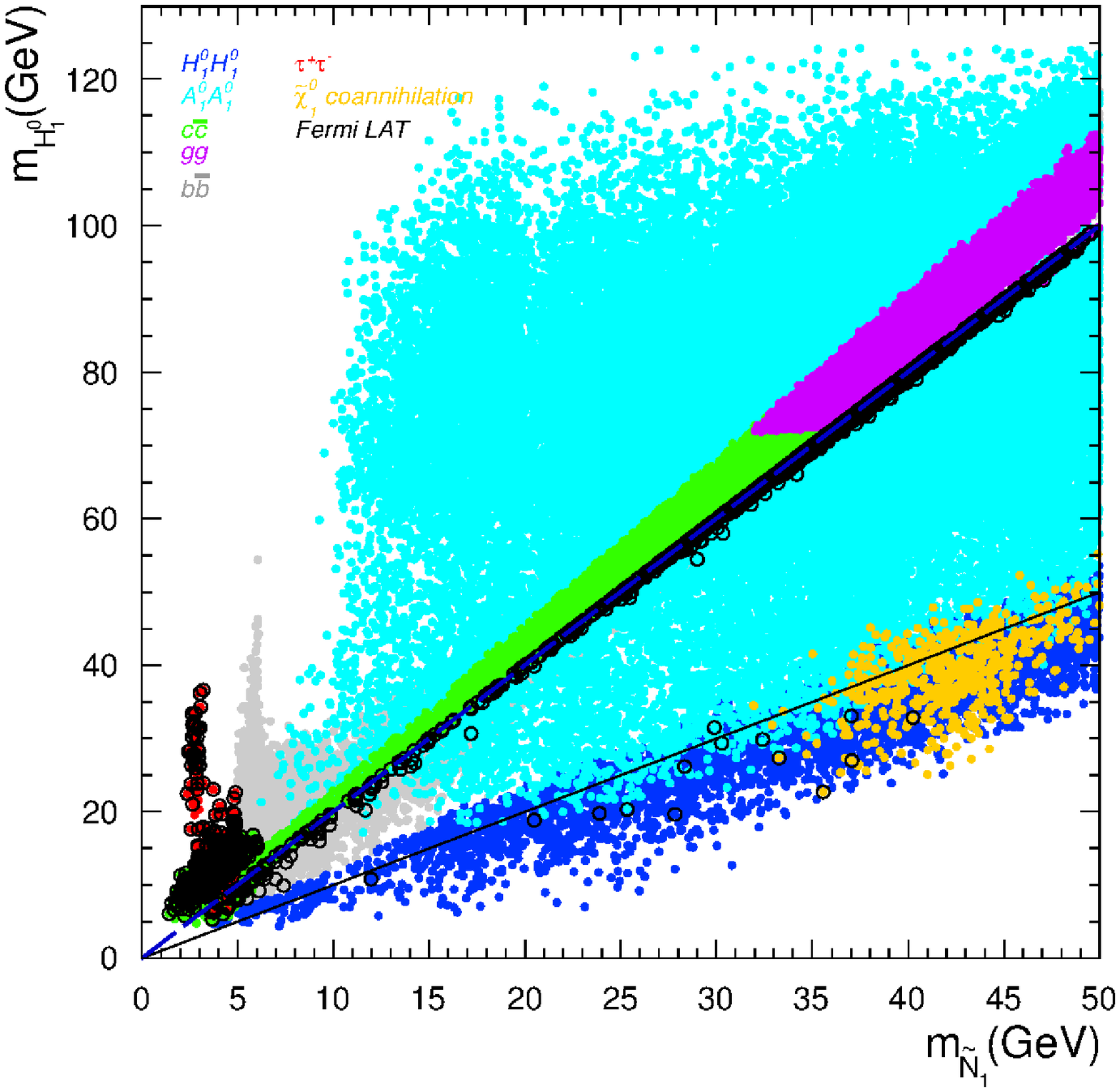,width=7.5cm}
	\end{center}
\caption{\small Lightest CP-even Higgs mass as a function of the RH sneutrino mass. All the points fulfil the experimental constraints and have a relic abundance $0.001<\Omega_{\tilde{N}_1}h^2<0.13$. The different colours indicate the dominant annihilation channel. 
The solid line corresponds to $m_{\tilde{N}_1}=m_{H^0_1}$ and the dashed line to $m_{\tilde{N}_1}=m_{H^0_1}/2$. 
The plot on the right-hand side incorporates the constraints from LUX, CDMSlite, and SuperCDMS 
on $\sigsi$. Black circles correspond to the points excluded by the Fermi LAT bounds from dSphs on the RH sneutrino annihilation cross-section. Points corresponding to different annihilation channels are shown separately in Fig.\,\ref{fig:mh1-s}.}
  \label{fig:mh1}
\end{figure}

\begin{figure}[t!]
	\begin{center}  
	\epsfig{file=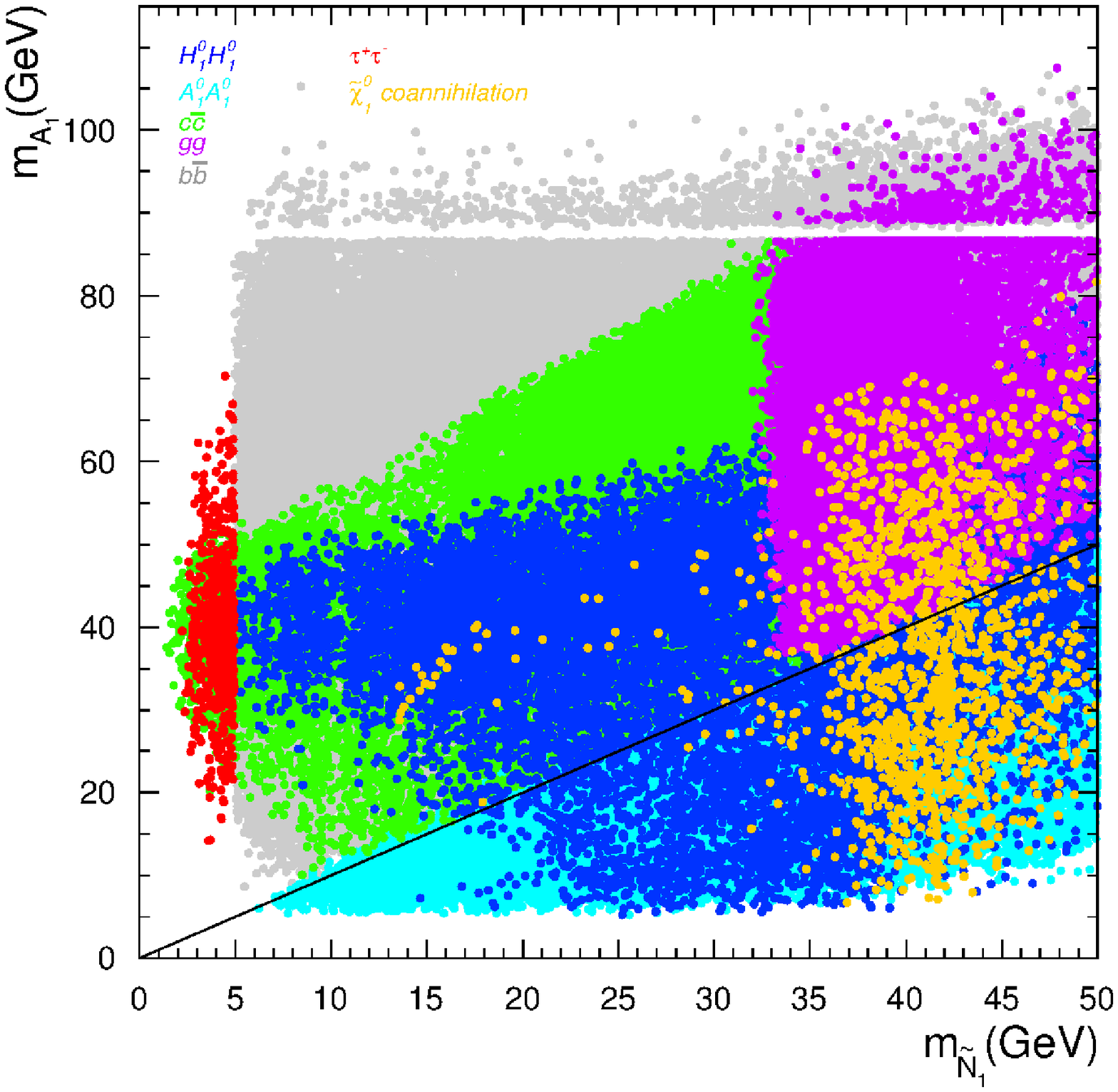,width=7.5cm}
	\epsfig{file=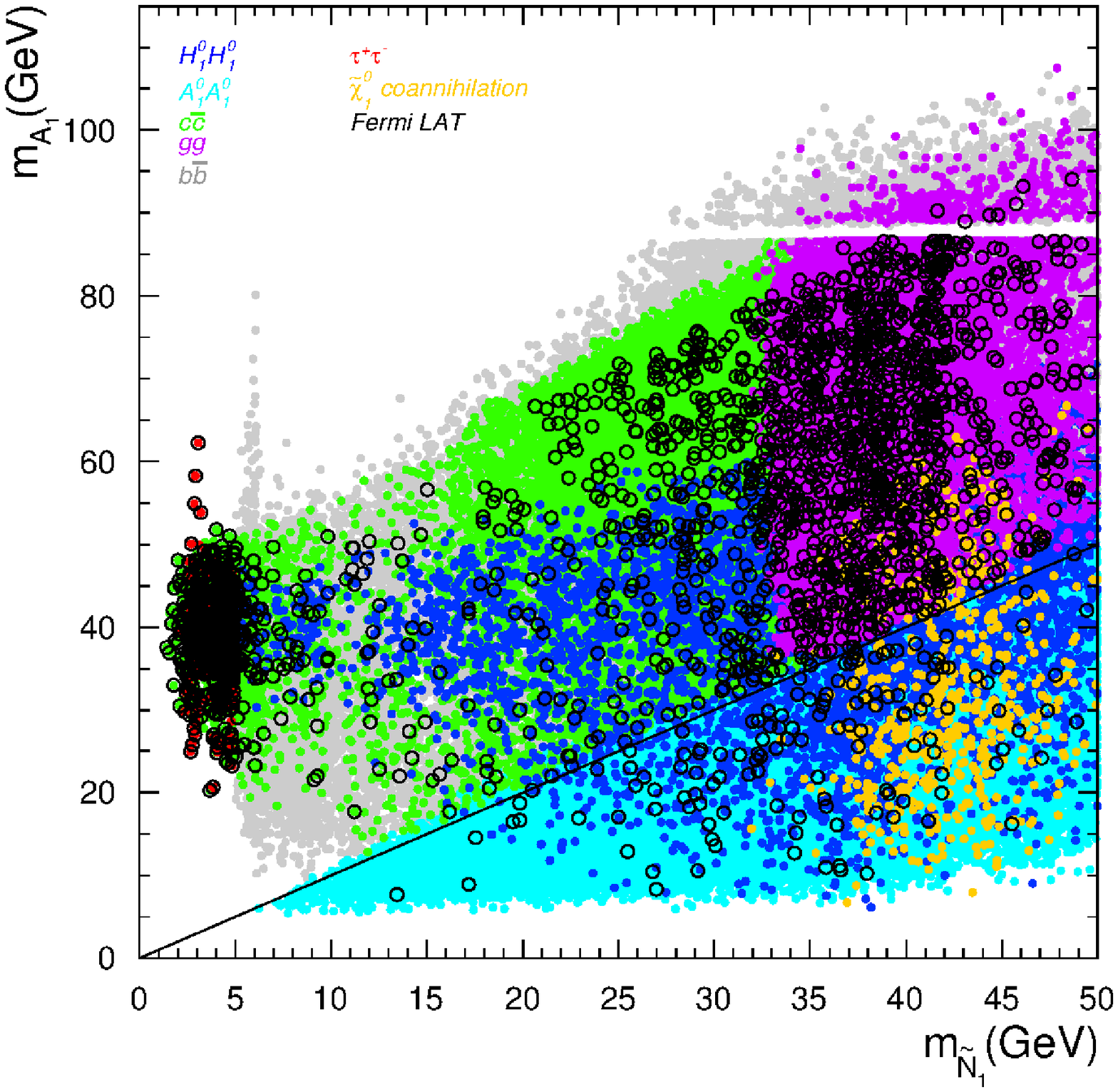,width=7.5cm}
	\end{center}
\caption{\small Lightest pseudoscalar Higgs mass as a function of the sneutrino mass for the dominant annihilation channels. 
The solid line corresponds to $m_{\tilde{N}_1}=m_{A^0_1}$. 
All points fulfil the experimental constraints and have a relic abundance $0.001<\Omega_{\tilde{N}_1}h^2<0.13$.
The plot on the right-hand side includes the constraints from LUX, CDMSlite, and SuperCDMS 
on $\sigsi$. Black circles represent the points excluded by the Fermi LAT bounds from dSphs on the RH sneutrino annihilation cross-section. Points corresponding to different annihilation channels are shown separately in Fig.\,\ref{fig:ma1-s}.}
  \label{fig:ma1}
\end{figure}

When the relic density and the Higgs sector constraints are applied to the scan in the parameter space of the model, a clear structure of the Higgs sector emerges, where the lightest Higgs boson has a mass $\hmassl\lesssim 120$~GeV and is singlet-like, whereas the SM Higgs corresponds to the second-lightest state and reproduces the experimental value for its mass, $\hmassm\approx126$~GeV.
On top of this, a light pseudoscalar Higgs is often found with $\phmassl\lesssim90$~GeV. 
In Fig.\,\ref{fig:mh1}, we have represented the mass of the lightest Higgs as a function of the RH sneutrino mass for the points that pass all the experimental constraints. The plot on the right-hand side of Fig.\,\ref{fig:mh1} also incorporates bounds on direct and indirect DM searches. 
Similarly, Fig.\,\ref{fig:ma1} displays the mass of the lightest pseudoscalar Higgs as a function of the RH sneutrino mass.

The colours in these plots indicate the various possible annihilation final states and point towards a rich phenomenology. We have identified the following possibilities,
\begin{itemize}
\item\textbf{$b\bar{b}$} (grey):  It is driven by a s-channel Higgs exchange and it is the most common annihilation final state 
for very light sneutrinos since the $b$-quark Yukawa coupling dominates over the other couplings.
\item\textbf{$\tau^{+}\tau^{-}$} (red): This channel is generally dominant when the $b\bar{b}$ final state is closed, i.e. when 
$m_{\tilde{N}_1}<m_b$.
\item\textbf{$c\bar{c}$} (green):  This final state prevails in those regions of the parameter space in which the lightest 
Higgs is predominantly singlet-like and its composition is such that $|S^{2}_{H^0_1}/S^{1}_{H^0_1}|\gtrsim5\tanb$.
\item\textbf{$gg$} (violet):  This channel is subject to the same conditions as the previous one, but additionally it requires
$m_{H^0_1}\gtrsim 80$~GeV. Otherwise, since it is driven by a top quark loop, it is suppressed with respect to the $c\bar{c}$ final state.
Both $c\bar{c}$ and $gg$ channels are mostly produced in resonant annihilations, with  
$m_{H^0_1}\simeq 2m_{\tilde{N}_1}$ (dashed line in Fig.~\ref{fig:mh1}).
\item\textbf{$H^0_1H^0_1$} (blue): In the NMSSM a very light CP-even Higgs is viable as long as  it is singlet-like.
When this channel is kinematically open, it usually dominates. This fact is well illustrated in Fig.~\ref{fig:mh1} 
where the $H^0_1H^0_1$ final states gather below the solid line that corresponds to $m_{\tilde{N}_1}=m_{H^0_1}$. 
\item\textbf{$A^0_1A^0_1$} (cyan): A very light CP-odd Higgs is also viable in the NMSSM, provided that
its singlet component is large as well. 
When this channel is kinematically allowed, it easily prevails unless the channel $H^0_1H^0_1$ is open.
\item Coannihilation with the lightest neutralino (orange): This happens when $\snmassr\approx\neutmass$. 
Notice, that the lightest neutralino can also be very light in the NMSSM if it is mostly bino or singlino. However, since it also contributes to the invisible decay of the SM Higgs, its couplings are constrained. In our scan, co-annihilation effects are important for a small population of points with $\snmassr\gsim30$~GeV for which the light neutralino is singlino-like. 
\end{itemize}

\begin{figure}[!t]
	\begin{center}  
	\epsfig{file=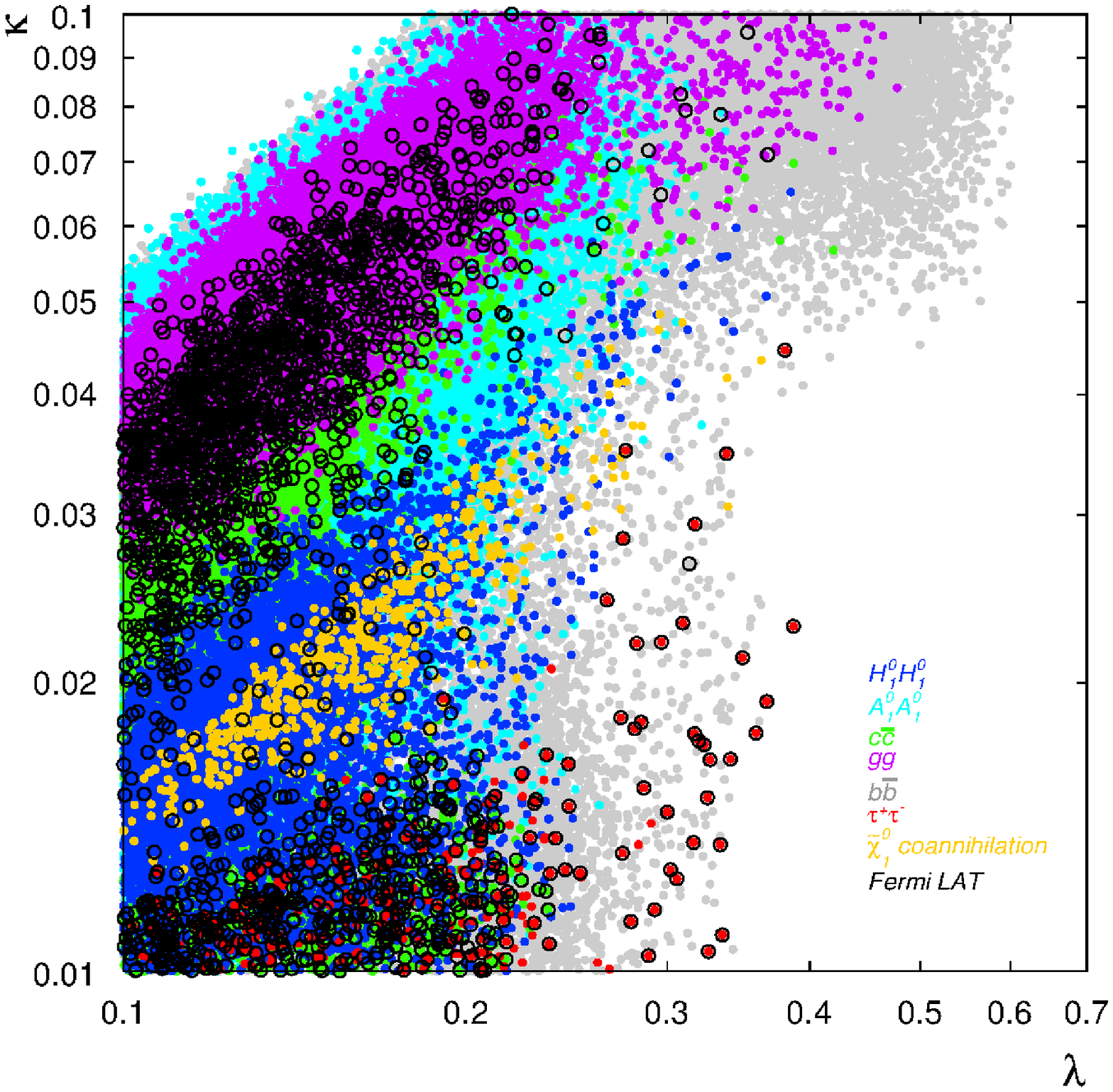,width=7.5cm}
	\epsfig{file=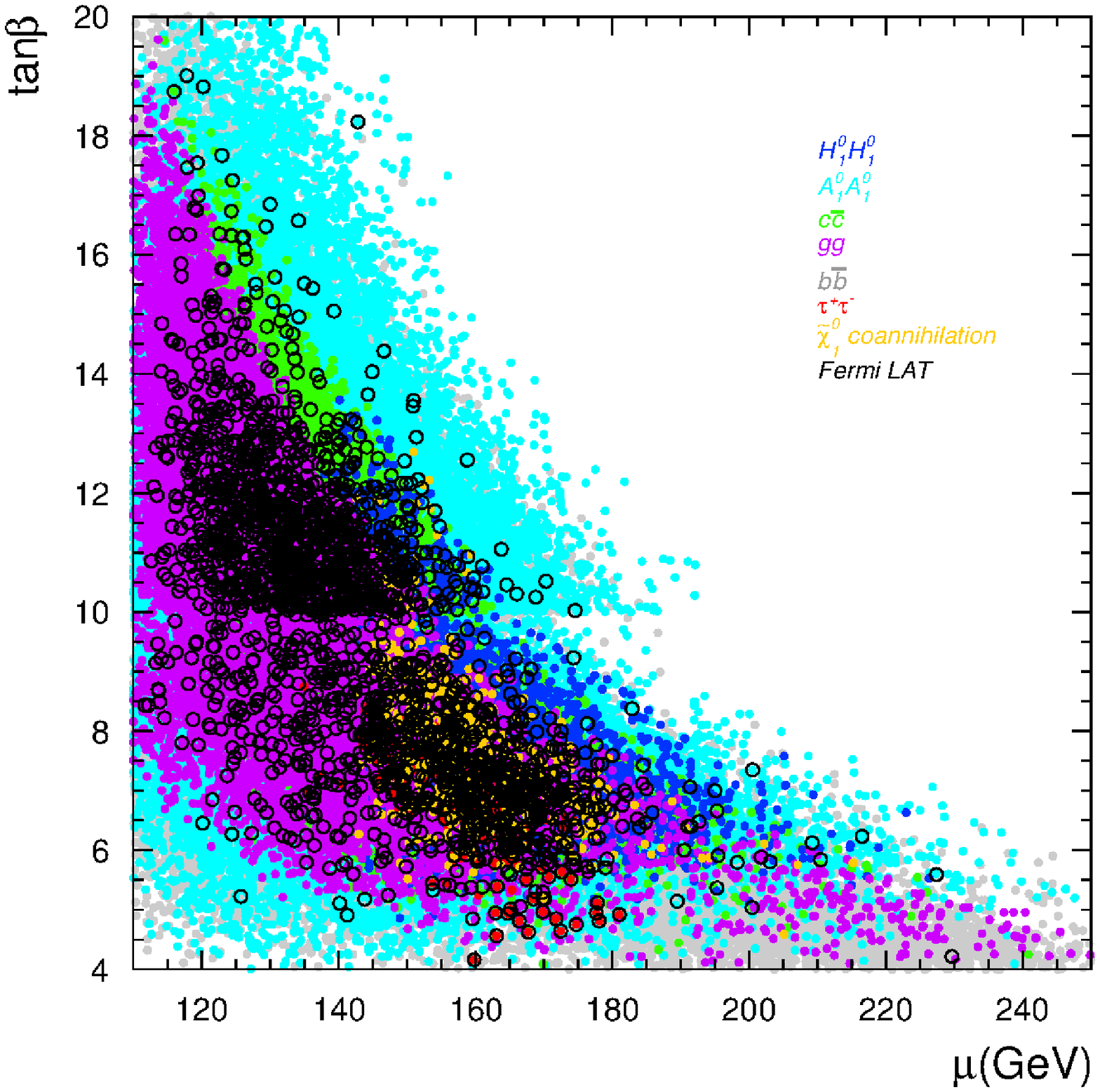,width=7.5cm}\\
	\epsfig{file=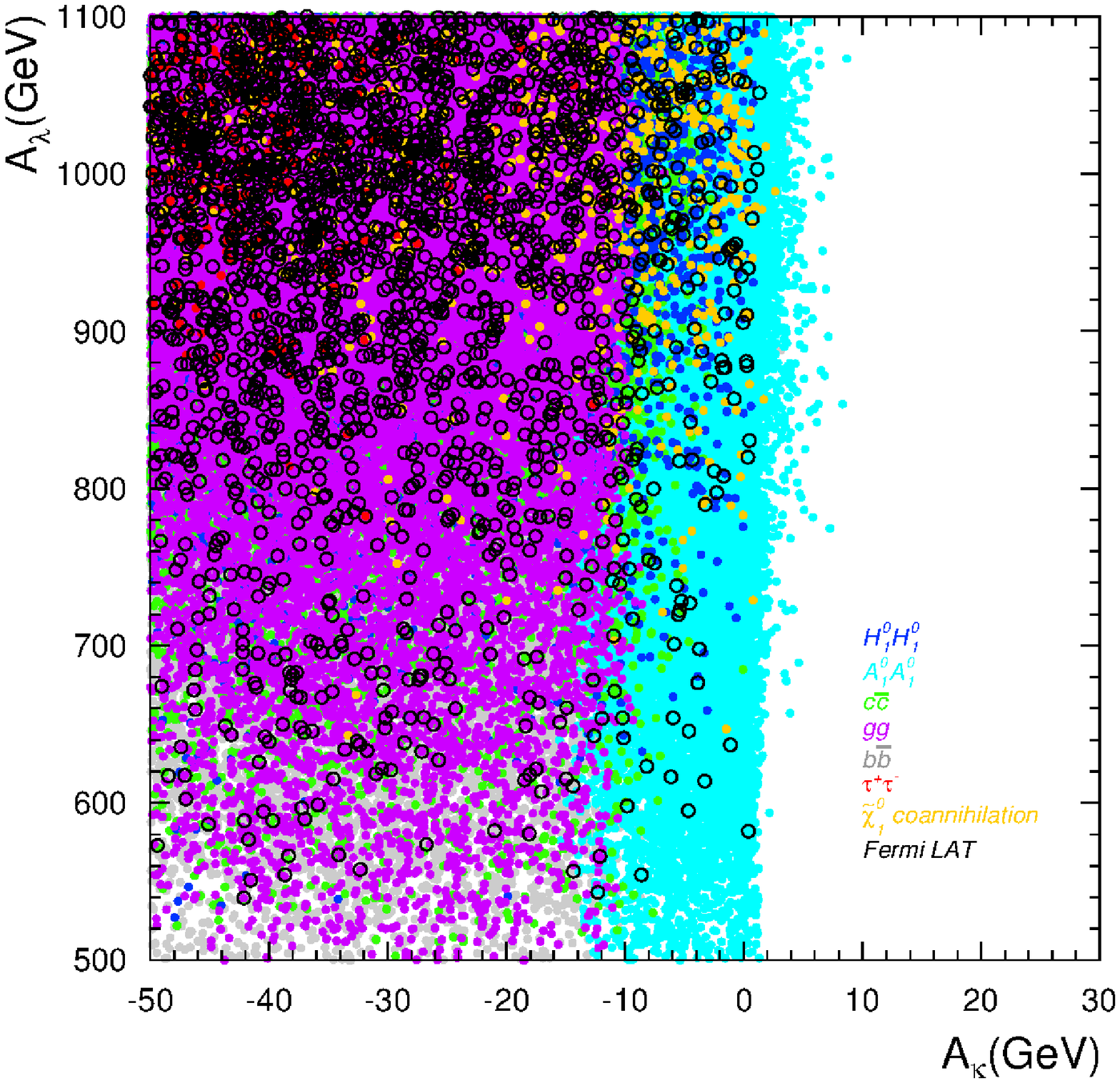,width=7.5cm}
	\epsfig{file=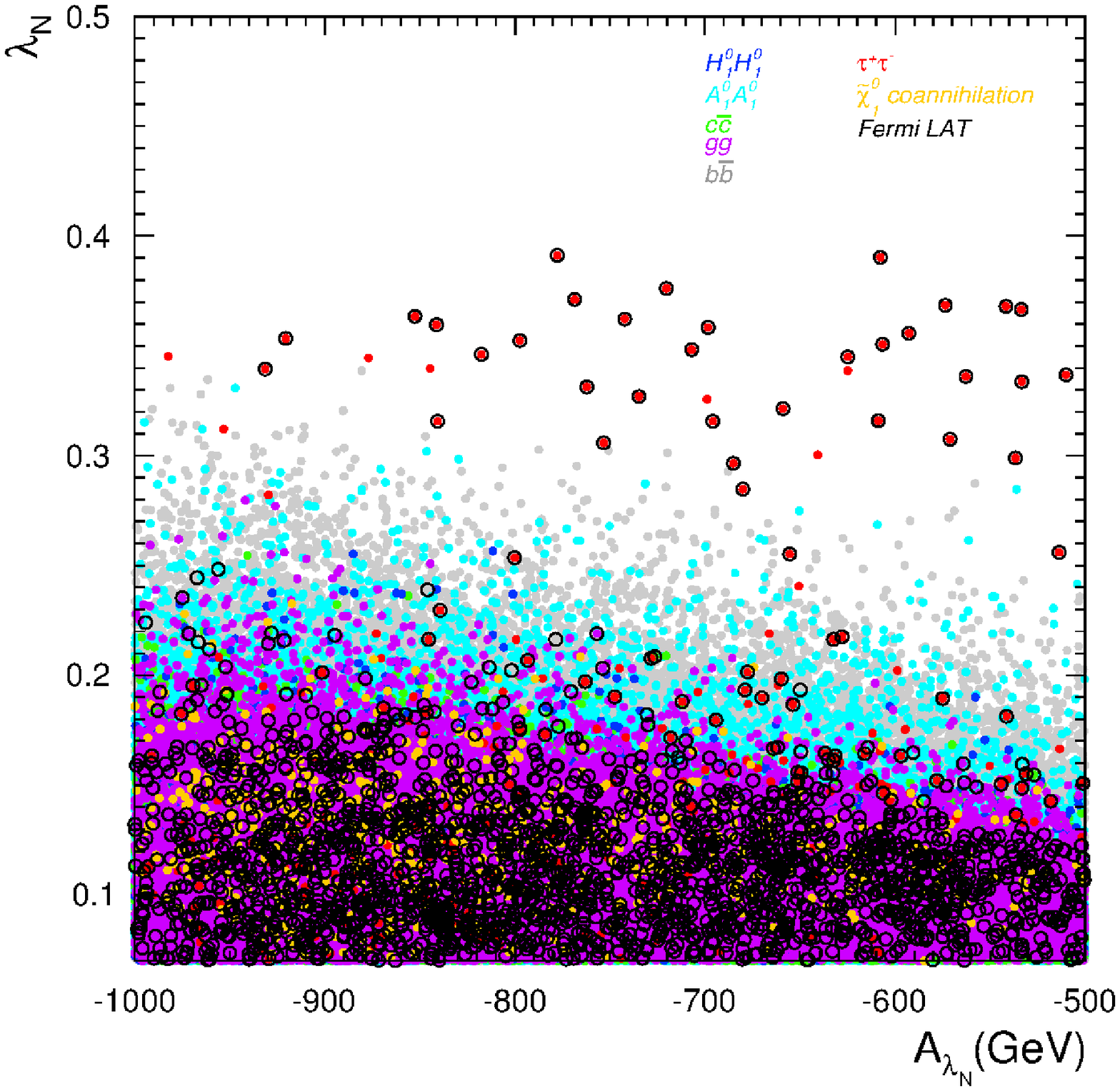,width=7.5cm}
	\end{center}
\caption{\small Scatter plot in the $(\lambda,\,\kappa)$, $(\mu\,\tan\beta)$, $(\ak,\,\al)$, and $(\ln\,\aln)$ planes from left to right and top to bottom.
All the experimental constraints have been included, as well as the constraints from LUX, CDMSlite, and SuperCDMS 
on $\sigsi$. Black circles correspond to the points excluded by the Fermi LAT bounds from dSphs on the RH sneutrino annihilation cross-section. 
The colour code is the same as in Fig.\,\ref{fig:mh1}.
 }
\label{fig:scan}	
\end{figure}

The plots on the right-hand side of Fig.\,\ref{fig:mh1} and Fig.\,\ref{fig:ma1} incorporate also the constraints from LUX, CDMSlite, SuperCDMS on $\sigsi$, i.e., points that do not fulfil this bound have been removed.
As we can observe, this has an important effect in some channels. In the case of $b\bar b$, most of the points with $\snmassr\gsim 7$~GeV are excluded and only those which satisfy the resonance condition with the lightest Higgs remain 
(direct detection bounds are less severe for masses below 7~GeV).
Other channels are less affected by these bounds, such as points in which the main annihilation channel is $A^0_1A^0_1$ and those with resonant annihilation ($c\bar c$ and $gg$). 
As we will argue in Section\,\ref{sec:direct}, the resulting RH sneutrino scattering cross-section for these points is small, below current direct detection bounds.

Finally, black circles on the right-hand side of Fig.\,\ref{fig:mh1} and Fig.\,\ref{fig:ma1} denote the points for which the RH sneutrino annihilation cross-section exceeds the upper constraints obtained from Fermi LAT observations of the gamma ray flux of dwarf spheroidal galaxies. 
This bound excludes points with  resonant annihilation and many of the points with low mass (but not all). 
We will study the effect of this limit in more detail in Section\,\ref{sec:indirect}.

In order to illustrate the effect of all the experimental bounds, including those from direct searches, on the initial parameter space, we show in Fig.\,\ref{fig:scan} the scatter plots in the 
$(\lambda,\,\kappa)$, $(\mu,\,\tan\beta)$, $(\ak,\,\al)$, and $(\aln,\,\ln)$ planes\footnote{The feature observed in the $(\mu,\,\tan\beta)$ plane for $\tanb=10$ is an artifact due to 
having split the scan in two ranges of $\tan\beta$.}.
These plots show that wide areas of the parameter space are viable.
Nevertheless, a slight preference for small values of $\lambda$ and, especially, $\kappa$, in order to obtain a light singlet-like CP-even Higgs, is observed. 
Small values of $\kappa$ and $|\ak|$ also lead to a light pseudoscalar Higgs and we can notice that $|\ak|<10$~GeV is necessary so that the $A^0_1A^0_1$ channel is open and dominant(cyan points).
Note that if $\tan\beta$ increases the $\mu$ parameter decreases.

\begin{figure}[!t]
	\begin{center}  
	\epsfig{file=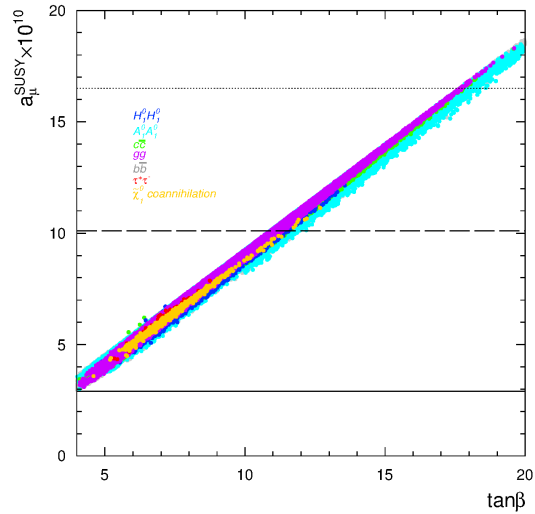,width=7.5cm}
	\end{center}
\caption{\small Theoretical predictions for the SUSY contribution to the muon anomalous magnetic moment as a function of $\tan\beta$. 
 The solid, dashed and dotted lines indicate the lower bounds at $2\sigma$ from tau data, $e^+e^-$ data, and the combined result using the HLS model, respectively. 
 All the experimental constraints, including those from direct and indirect DM searches, have been applied and the colour code is as in Fig.\,\ref{fig:mh1}. }
\label{fig:g-2}	
\end{figure}

Finally, in Fig.\,\ref{fig:g-2} we represent the theoretical predictions for $\asusy$ as a function of $\tan\beta$ for the points passing all the experimental constraints, including also those from indirect and direct dark matter searches.  
Being this an observable that favours low mass SUSY, its implications are in general in tension with the predictions of SUSY models when the bound on the Higgs mass is imposed, 
as it has been pointed out in previous analyses  \cite{Endo:2011gy,Endo:2013bba}.
However, in our construction we can observe that the predicted contribution to the SM value, despite small, is consistent with the current experimental constraints from tau data, and for $\tan\beta\gsim 10$ also with the results from $e^+e^-$. 
Compatibility with the region obtained using the HLS model can be achieved for $\tan\beta\gsim 16$.

\subsection{Indirect Detection}
\label{sec:indirect}

The analysis of the gamma ray spectrum from the Galactic Centre and from dwarf spheroidal galaxies by the Fermi LAT collaboration has shown no clear signal that can be attributed to dark matter.
These analyses have been used to extract upper constraints on the DM pair annihilation cross-section in pure channels that can be interpreted as a lower bound on thermal DM candidates \cite{GeringerSameth:2011iw,Ackermann:2011wa,Gomez-Vargas:2013bea,Ackermann:2013yva} (i.e., those for which the annihilation cross-section in the Early Universe is 
$\langle\sigma v\rangle \sim 3\times 10^{-26}$ cm$^3$s$^{-1}$).
These constraints are channel dependent, for example, data from dSph galaxies implies $m_{DM}\gtrsim11.3$~GeV for the $b\bar b$ channel \cite{Ackermann:2013yva}. 
In the case of the GC, the extracted bounds are comparable if the emission from known point sources and from the Galactic disk is subtracted  
\cite{Hooper:2011ti,Hooper:2012sr}. Nevertheless, without assuming a background model, the bounds are significantly less stringent unless compressed DM profiles are considered \cite{Gomez-Vargas:2013bea}. 
Finally, limits have also been derived from
an analysis of the Andromeda galaxy \cite{Li:2013qya}
and studies of the diffuse gamma ray emission in our Galaxy \cite{Abdo:2010dk,Ackermann:2012rg,Ackermann:2012qk,fortheFermiLAT:2013naa,Tavakoli:2013zva}.
We will consider only the bounds from dSphs data obtained in Ref.~\cite{Ackermann:2013yva}, since these are currently the strongest ones obtained by the Fermi LAT collaboration.

It should be emphasised that when uncertainties associated with the dark matter density profiles of dSph galaxies are included, the resulting constraints on $\langle\sigma v\rangle_0$ can vary up to an order of magnitude 
\cite{GeringerSameth:2011iw}. For this reason, even though in our analysis we apply this bound at face value, we have preferred to 
indicate explicitly the disfavoured points instead of removing them from our plots.

\begin{figure}[t!]
	\begin{center}  
	\epsfig{file=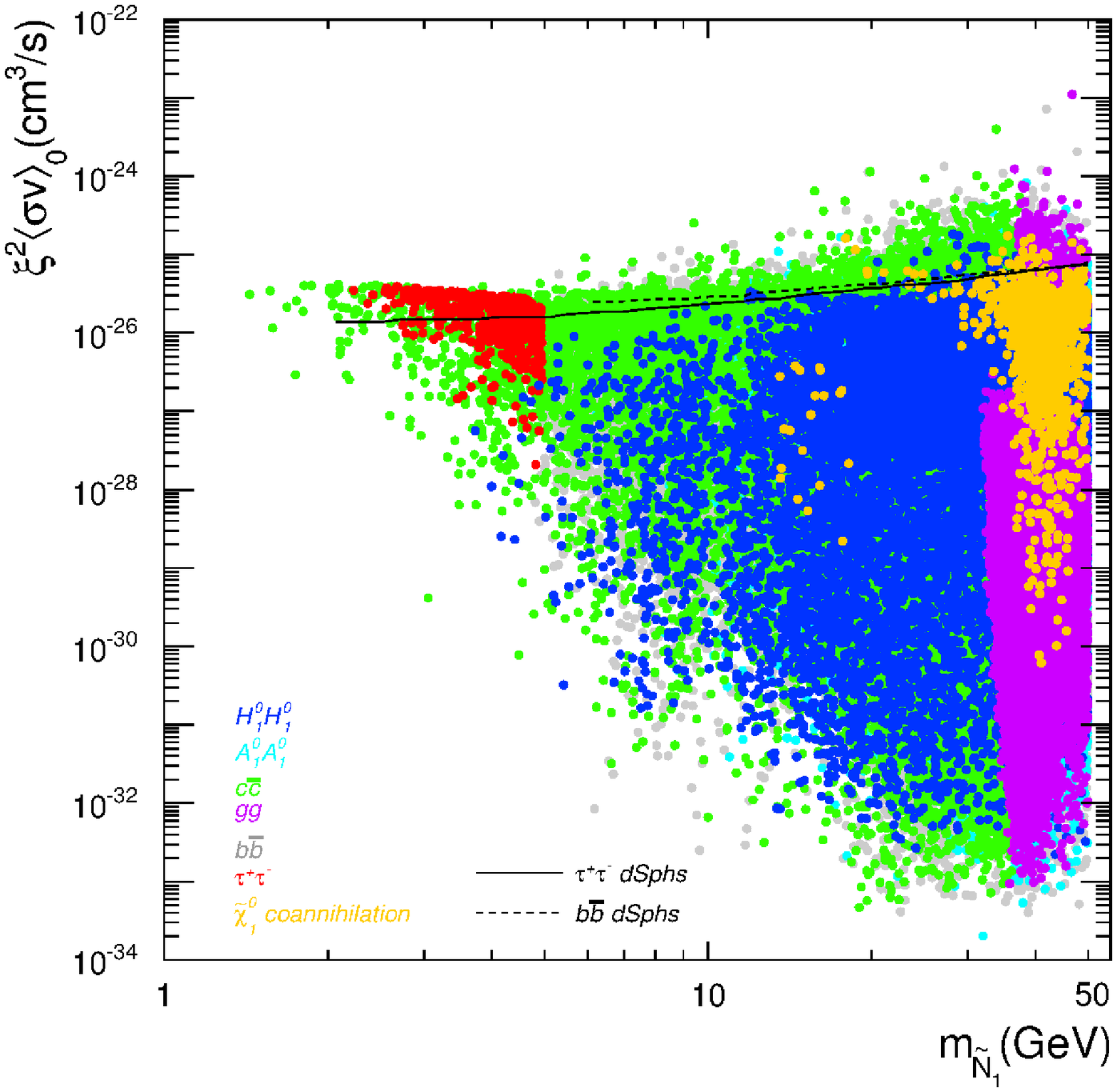,width=7.5cm}
	\epsfig{file=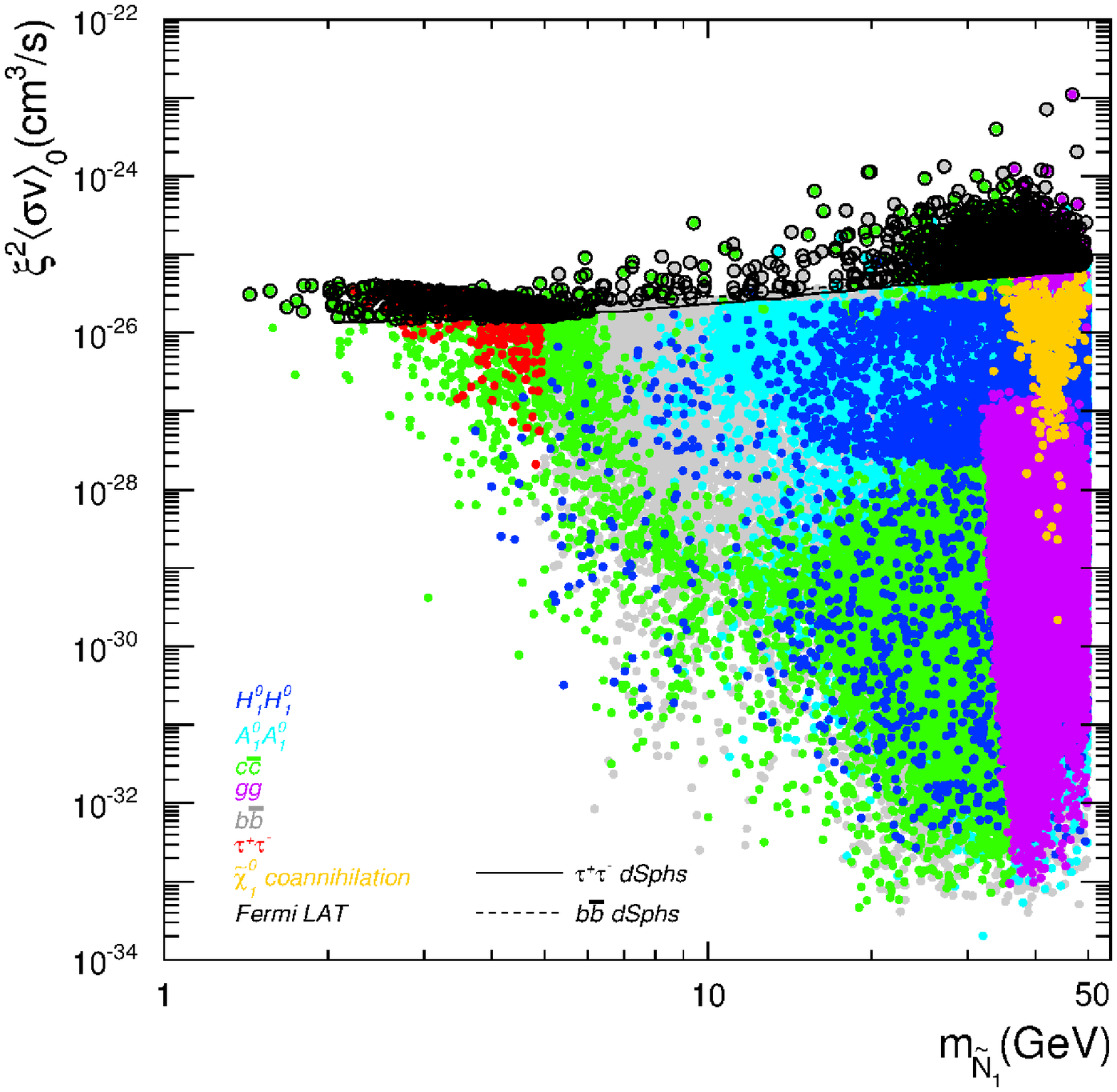,width=7.5cm}
	\end{center}
\caption{\small Thermally averaged RH sneutrino annihilation cross-section in the galactic halo as a 
function of the RH sneutrino mass. 
All the points fulfil the experimental constraints and have a relic abundance $0.001<\Omega_{\tilde{N}_1}h^2<0.13$. The solid and dashed lines correspond to the upper bounds on $\langle\sigma v\rangle_0$ derived from an analysis of dSph galaxies for pure $\tau^-\tau^+$ and $b\bar b$ channels, respectively. The plot on the right-hand side incorporates the constraints from LUX, CDMSlite, and SuperCDMS on $\sigsi$. The colour code is as in Fig.\,\ref{fig:mh1}.  Points corresponding to different annihilation channels are shown separately in Fig.\,\ref{fig:sigmav-s}.}
  \label{fig:sigmav}
\end{figure}

\begin{figure}[t!]
	\begin{center}  
	\hspace*{-0.5cm}\epsfig{file=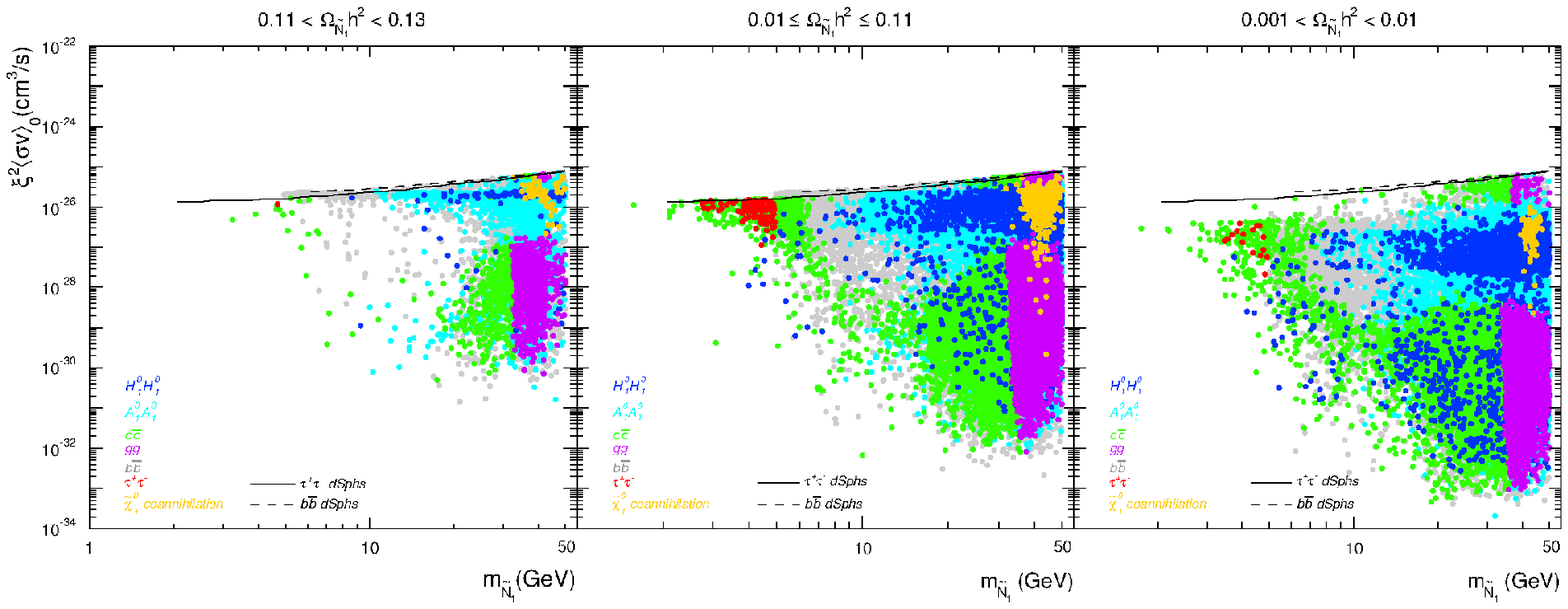,width=16cm}
	\end{center}
\caption{\small Thermally averaged RH sneutrino annihilation cross-section in the galactic halo as a 
function of the RH sneutrino mass.  From left to right, the three panels show different ranges in the RH sneutrino relic abundance, the leftmost one being in agreement with Planck results.
All the experimental constraints are included, together with the bounds from direct detection experiments and Fermi LAT data on dSphs. The colour code is as in Fig.\,\ref{fig:mh1}. 
}
  \label{fig:sigmav_relic}
\end{figure}

On the other hand, antimatter searches in cosmic rays are also suitable to constrain DM annihilation in the galactic halo.
Measurements of the antiproton flux performed by the PAMELA satellite \cite{Adriani:2010rc,Adriani:2012paa} agree very well with 
the expected astrophysical background. Consequently, these observations can be used to set bounds on the annihilation cross-section of 
light dark matter in hadronic channels. The resulting limits are comparable to those from Fermi LAT on dSphs, although $b\bar b$ and 
$c\bar c$ final states are especially constrained below $40$~GeV for the most probable set of parameters \cite{Fornengo:2013xda}. 
Even if these results are more stringent, they
are extremely sensitive to the model that describes the propagation of antiprotons in the Galaxy. In fact, if we take 
into account the uncertainty in the parameters describing these models, the bounds can be modified up to about an order of magnitude so that  
the $b\bar b$ and $c\bar c$ channels might move from a non-constrained region to one with stringent bounds.  

Lastly, searches for spectral features in the AMS positron fraction data\cite{Aguilar:2013qda} have been used to derive upper limits on $\langle\sigma v\rangle_0$ 
for leptonic channels \cite{Bergstrom:2013jra}. Since the bound on the $\tau^+\tau^-$ channel is placed above $\sim20$~GeV, 
these findings do not exclude any point in our data set.

In Fig.~\ref{fig:sigmav}, we show the thermally averaged RH sneutrino annihilation cross-section in the galactic halo, 
$\xi^2\langle\sigma v\rangle_0$ as a function of the RH sneutrino mass, where $\xi$ is the fractional density previously defined.
The entire set of points fulfil all the experimental constraints and reproduce the right relic abundance. 
The plot on the right-hand side also incorporates the limits from direct detection experiments.

As in the previous figures, black circles indicate the points for which the RH sneutrino annihilation cross-section exceeds the Fermi LAT constraint. It should be noted that upper bounds on $\langle\sigma v\rangle_0$ are derived for pure channels. Although in our analysis we do not obtain pure annihilation channels (but a mixture of various), we have implemented the corresponding constraint when the contribution of a specific channel was dominant over the others. 
For the $\higgsl\higgsl$, $\phiggsl\phiggsl$, and $gg$ channels we consider the same limit as for $b\bar b$. 
On the other hand, for a 10 GeV DM particle, the primary gamma ray spectra for the $\tau^+\tau^-$ and $c\bar c$ final states above 1 GeV are very similar \cite{Cirelli:2010xx} and the inverse Compton contribution is negligible for dark mater masses below 500 GeV in the energy range observed by the Fermi satellite \cite{Gomez-Vargas:2013bea}. 
Consequently, at first order we will apply the bounds for $\tau^+\tau^-$ to the $c\bar c$ final states.

As it is illustrated in Fig.~\ref{fig:sigmav}, Fermi LAT constraints compromise a small region of the allowed parameter space, but its effect is particularly severe for light masses. In particular, many of the examples with $\tau^+\tau^-$ and $c\bar c$ final states become disfavoured for this reason.
In some cases, the predicted flux is several orders of magnitude below the current experimental sensitivity. 
These points correspond to those in which RH sneutrino annihilation proceeds through a resonant s-channel Higgs-mediated 
diagram $m_{\snr_1}\simeq m_{H^0_1}/2$ or in which co-annihilation effects help to reduce the relic density. 
In these cases, the correlation between the annihilation cross-section in the Early Universe and in the DM halo is lost and the latter can be significantly smaller.

In Fig.~\ref{fig:sigmav_relic}, we have separated the results for $\xi^2\langle\sigma v\rangle_0$ according to the relic abundance of RH sneutrinos, displaying on the left panel the points in agreement with Planck results and in the other two panels examples in which the RH sneutrino would constitute only a fraction of the DM. The correct relic density can be obtained for most of the mass range (above $\snmassr\gsim 3$~GeV), although many of the examples with light masses (mainly those with predominant annihilation into $\tau^+\tau^-$ and $c\bar c$) tend to predict a smaller $\Omega_{\snr_1} h^2$.
On the other hand, it is noteworthy that there are numerous examples in which RH sneutrinos could be observable despite being underabundant.

Finally, notice that if the low-energy excess in the gamma ray flux from the Galactic Centre was confirmed, the analyses of Refs.\,\cite{Vitale:2009hr,Morselli:2010ty,Hooper:2010mq,Hooper:2011ti,Abazajian:2012pn,Hooper:2013nhl,Daylan:2014rsa} suggest that it could be explained by annihilation of light dark matter particles.
The analysis of this excess in Ref.\,\cite{Gordon:2013vta} favours DM with a mass in the range of around $8-65$~GeV, annihilating preferentially to a mixture of $\tau^{+}\tau^{-}$ and $b\bar{b}$ final states with  $\langle\sigma v\rangle_0$ of order 
of the thermal relic value (see Fig.~10 and Table.~IV in Ref.\,\cite{Gordon:2013vta}). Several particle physics solutions have been proposed in this mass range~\cite{Belikov:2010yi,Buckley:2010ve,Zhu:2011dz,Marshall:2011mm,Buckley:2011vs,Boucenna:2011hy,Buckley:2011mm,Hooper:2012cw,Cotta:2013jna,Buckley:2013sca,Hagiwara:2013qya,Fortes:2013ysa,Alves:2013tqa,Modak:2013jya,Boehm:2014hva}. 
In Table~\ref{tab:fermiGC}, we show some examples of the RH sneutrino parameter space with dominant $b\bar{b}$ annihilation inside the $5\sigma$ confidence region derived in Ref.\,\cite{Gordon:2013vta}, 
as well as an example for the best-fit point.
Findings in Ref. \cite{Gordon:2013vta} are in agreement with the results of a recent work, which favours $31-40$~GeV DM annihilating to $b\bar b$ with $\langle\sigma v\rangle_0 = (1.4 - 2.0)\times10^{-26}$ cm$^3$s$^{-1}$ \cite{Daylan:2014rsa}.
We also provide an example compatible with this range.

\begin{table}[!t]
  \begin{center}
    \begin{tabular}{|c|c|cccc|c|c|}
      \hline
      $\snmassr$(GeV) & $\xi^2\langle\sigma v\rangle_0$(cm$^3$/s) & $\%b\bar{b}$ & $\%\tau^{+}\tau^{-}$ & $\%c\bar{c}$ & $\%gg$ & $\xi\sigsi$(pb) & $\Omega_{\snr_1} h^2$ \\
      \hline
      $11.34$ & $1.38\times 10^{-26}$ & $88.39$ & $6.83$ & $4.07$ & $0.67$ & $2.38\times 10^{-11}$ & $0.0696$ \\      
      $12.80$ & $1.21\times 10^{-26}$ & $88.88$ & $6.72$ & $3.51$ & $0.83$ & $5.97\times 10^{-10}$ & $0.1126$\\
      $13.32$ & $1.34\times 10^{-26}$ & $86.72$ & $6.68$ & $5.61$ & $0.92$ & $4.12\times 10^{-10}$ & $0.0374$ \\      
      $14.49$ & $1.31\times 10^{-26}$ & $78.57$ & $6.18$ & $13.02$ & $2.11$ & $1.29\times 10^{-9}$ & $0.0201$\\      
      $17.78$ & $2.18\times 10^{-26}$ & $84.22$ & $6.62$ & $7.08$ & $1.98$ & $4.19\times 10^{-11}$ & $0.0597$\\      
      $21.50$ & $1.76\times 10^{-26}$ & $86.34$ & $6.91$ & $4.68$ & $1.98$ & $2.96\times 10^{-10}$ & $0.0634$\\      
      $23.60^*$\hspace*{-1ex} & $2.10\times 10^{-26}$ & $86.35$ & $7.05$ & $4.25$ & $2.19$ & $7.28\times 10^{-10}$ & $0.0717$\\  
      $27.71$ & $1.67\times 10^{-26}$ & $84.61$ & $7.06$ & $4.76$ & $3.43$ & $3.60\times 10^{-13}$ & $0.0366$\\
      $29.60$ & $1.95\times 10^{-26}$ & $86.30$ & $7.32$ & $3.32$ & $2.93$ & $3.34\times 10^{-14}$ & $0.0287$\\
      $34.79$**\hspace*{-2ex} &$1.74\times10^{-26}$& $89.97$& $7.58$& $0.78$ & $1.59$ & $3.49\times10^{-11}$ & $0.0613$\\       
      $42.96$ & $3.10\times 10^{-26}$ & $89.86$ & $7.96$ & $0.58$ & $1.50$ & $7.07\times 10^{-10}$ & $0.1045$\\      
      $43.96$ & $3.19\times 10^{-26}$ & $91.29$ & $8.03$ & $0.00$ & $0.62$ & $2.92\times 10^{-11}$ & $0.1137$\\      
      $45.68$ & $3.08\times 10^{-26}$ & $68.79$ & $5.79$ & $10.30$ & $13.97$ & $3.75\times 10^{-12}$ & $0.0068$ \\
      $48.26$ & $3.11\times 10^{-26}$ & $89.01$ & $7.86$ & $1.07$ & $1.53$ & $5.27\times 10^{-10}$ & $0.1219$\\      
      \hline
    \end{tabular}
    \vspace{0.4cm}
    \caption{\small Sample points within the $5\sigma$ region consistent with the observed low-energy excess in the 
     gamma-ray emission at the GC identified in the Fermi LAT data (see Fig.~10 and Table.~IV in Ref.~\cite{Gordon:2013vta}). For each point we indicate the
    RH sneutrino mass, thermally averaged annihilation cross-section, annihilation final states, and spin-independent elastic-scattering cross-section.
      The asterisk denotes an example compatible with the best fit point of Ref.~\cite{Gordon:2013vta} and two asterisks correspond to an example for Ref.\,\cite{Daylan:2014rsa}.
     }
    \label{tab:fermiGC}
  \end{center}
\end{table}

\subsection{Direct Detection}
\label{sec:direct}

Let us finally address the prospects for the direct detection of very light RH sneutrino dark matter. 
The interaction between RH sneutrinos and quarks can be described in terms of an effective Lagrangian, valid in the 
non-relativistic regime where collision takes place.  
In this case, there is only one Feynman diagram contributing at tree-level to this process, namely, 
the $t$-channel exchange of neutral Higgses, leading to a  scalar coupling, 
\begin{equation} 
  {\cal L}_{eff}\supset \alpha_{q_i} \snr_1\snr_1  \bar q_i q_i\ ,
\end{equation}
with
\begin{equation}
  \alpha_{q_i}\equiv\sum_{j=1}^3\frac{\chsnrsnr Y_{q_i}}{\hmassj^2}\ ,
  \label{alphaq}
\end{equation}
where $\chsnrsnr$ is the sneutrino-sneutrino-Higgs coupling \cite{Cerdeno:2009dv}, $Y_{q_i}$ is the
corresponding quark Yukawa coupling, and $i$ labels up-type quarks ($i=1$) and down-type quarks ($i=2$). 
The resulting spin-independent part of the sneutrino-nucleon elastic scattering cross-section reads
\begin{equation}
\sigsi=\frac{f_p^2m_p^2}{4\pi(\snmassr+m_p)^2}\ , 
\label{eq:sigsi}
\end{equation}
where $m_p$ is the proton mass, and $f_p$ is defined in terms of the hadronic matrix elements $f_{Tq_i}^p$ and $f_{TGP}^p$,
\begin{equation}
  \frac{f_p}{m_p}=
  \sum_{q_i=u,d,s}f_{Tq_i}^p\frac{\alpha_{q_i}}{m_{q_i}}+ 
  \frac{2}{27}\ f_{TG}^p\sum_{q_i=c,b,t}\frac{\alpha_{q_i}}{m_{q_i}}\ .
  \label{fpsneutrino}
\end{equation}
Expression (\ref{eq:sigsi}) corrects a missing factor in Refs. \cite{Cerdeno:2008ep,Cerdeno:2009dv,Cerdeno:2011qv}. 
Since the RH sneutrino is a scalar field, there is no contribution to the spin-dependent cross-section.

\begin{figure}[t!]
	\begin{center}  
	\epsfig{file=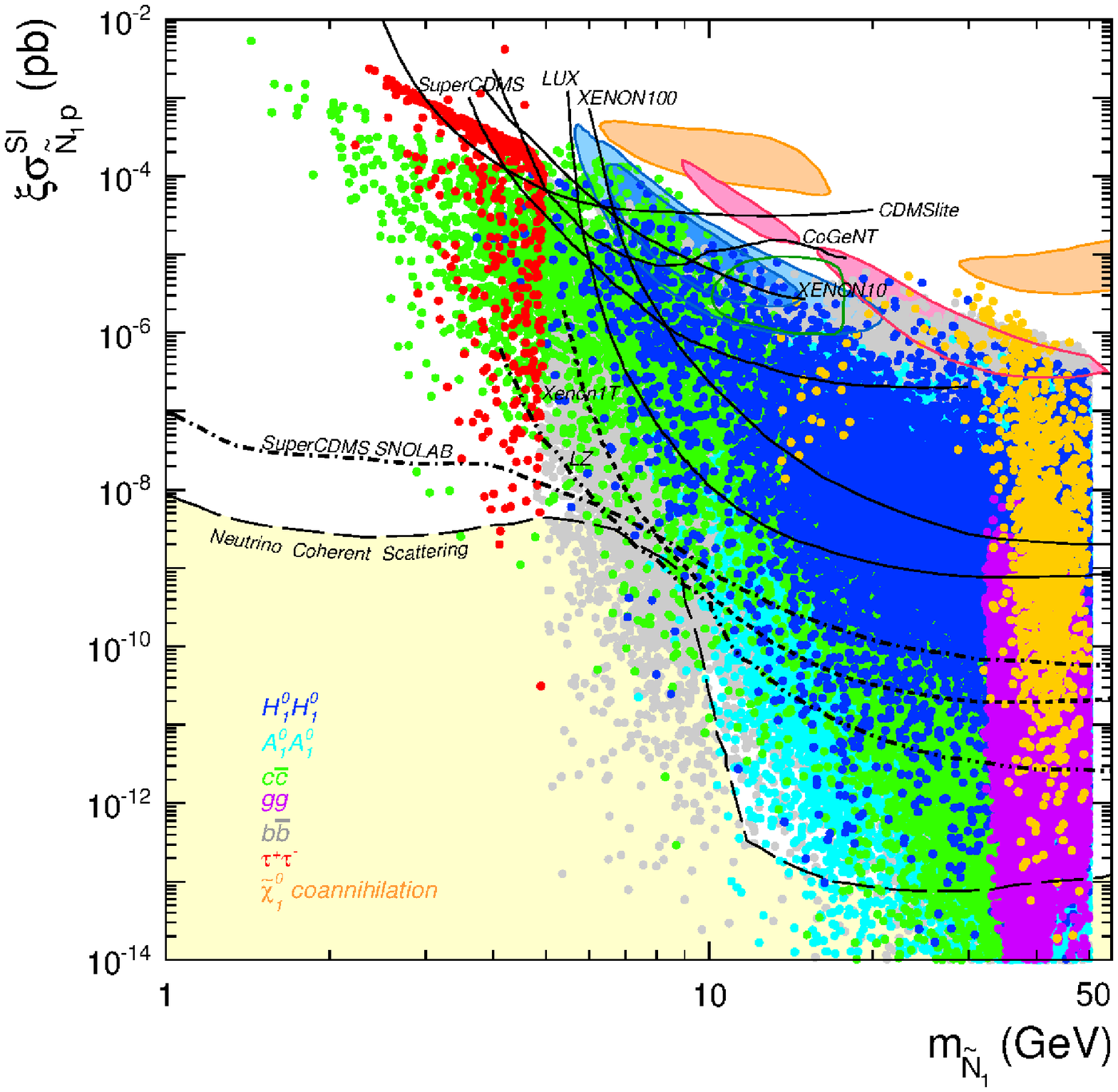,width=7.5cm}
	\epsfig{file=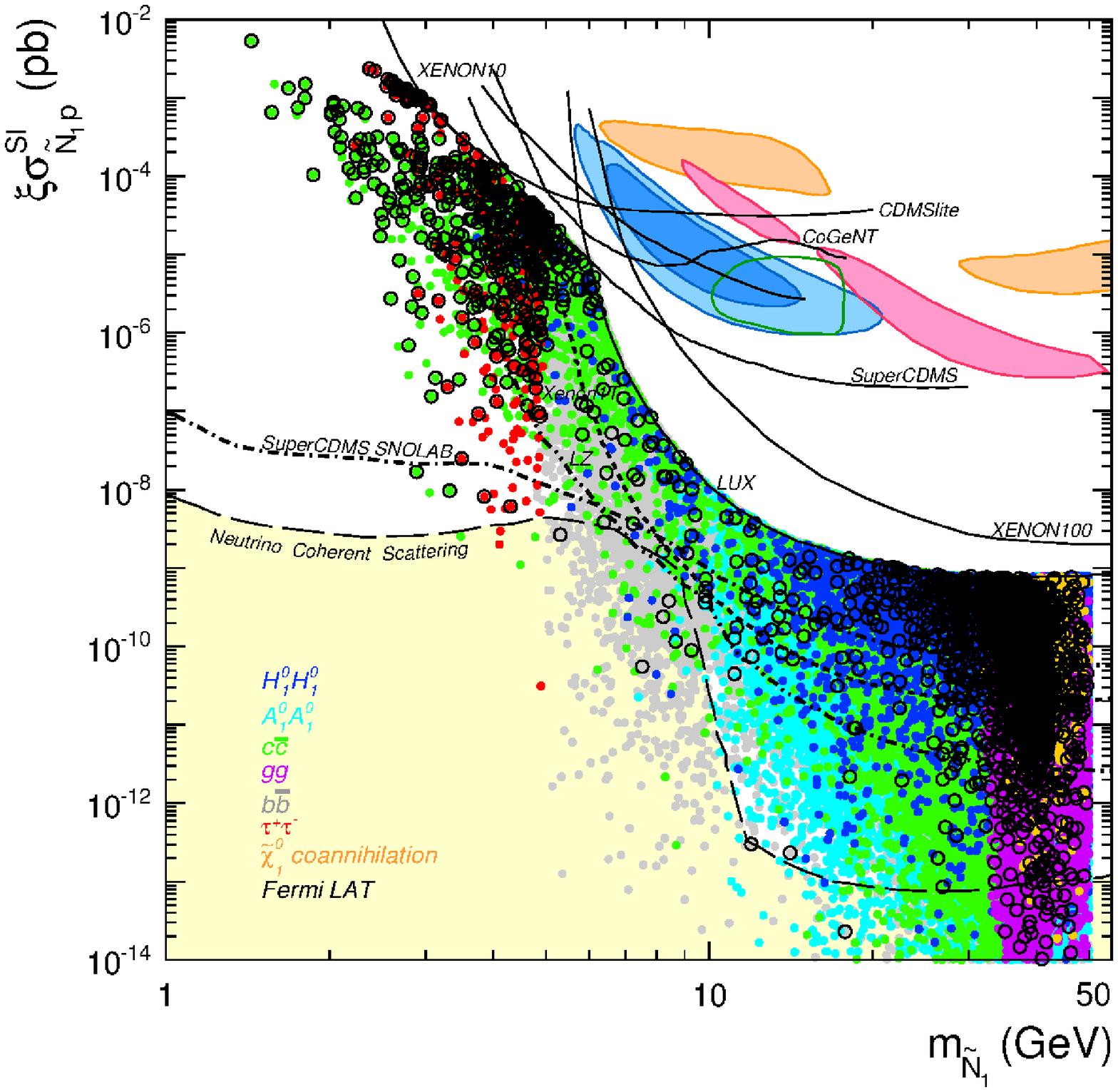,width=7.5cm}
	\end{center}
\caption{\small Theoretical predictions for $\sigsi$ as a function of the RH sneutrino mass. Solid lines represent the current  experimental upper bounds from direct detection experiments, whereas dotted lines are the projected sensitivities of next-generation detectors. 
The dashed line corresponds to an approximate band where neutrino coherent scattering with nuclei will begin to limit the sensitivity of direct detection experiments. 
Closed contours represent the areas compatible with the observed excesses in  DAMA/LIBRA (orange), CRESST (red), CDMS II (blue),  and CoGeNT (green).
The plot on the right-hand side includes the upper constraints from LUX, CDMSlite and SuperCDMS on $\sigsi$. Black circles correspond to points for which the annihilation cross-section exceeds the Fermi LAT bounds on dSph galaxies.  Points corresponding to different annihilation channels are shown separately in Fig.\,\ref{fig:spin-independent-s}.
}
  \label{fig:spin-independent}
\end{figure}

The theoretical predictions for $\xi\sigsi$ as a function of the RH sneutrino mass are depicted in  Fig.~\ref{fig:spin-independent}. As it was pointed out in Section\,\ref{sec:constraints}, $\xi$ is the RH sneutrino fractional density in the Galactic halo. 
We also show the most recent experimental upper constraints (solid lines) on the spin-independent scattering cross-section, as well as the regions compatible with the various hints for low-mass WIMPs. For reference, the predicted sensitivities of future experiments are also displayed. 
Finally, the dashed line represents the approximate band where neutrino coherent scattering with nuclei will begin to limit the sensitivity of direct detection experiments \cite{Billard:2013qya}.

In the plot on the left-hand side we have included all the experimental constraints detailed in Section\,\ref{sec:constraints}, 
except the direct and indirect detection bounds. 
The predicted $\sigsi$ spans many orders of magnitude for the whole range of RH sneutrino masses. Notice that it is possible to obtain points within the areas compatible with the hints for low mass WIMPs, and in particular within the CDMS II (Si) or CoGeNT regions\footnote{Of course, since the RH sneutrino is a standard WIMP, this does not solve the question of why they would have not been observed by XENON, LUX or SuperCDMS. }.

On the right-hand side of Fig.~\ref{fig:spin-independent}, we impose at face value the upper bounds on $\sigsi$ from direct detection experiments\footnote{Strictly speaking, these bounds correspond to the standard halo model with a local dark matter density of $\rho=0.3$~GeV cm$^{-3}$, a escape 
velocity of $v_{esc}=544$~km s$^{-1}$ , and a central velocity $v_0=220$~km s$^{-1}$. It is well known that deviations from this model can lead to shifts in the excluded regions. 
}
(LUX \cite{Akerib:2013tjd}, CDMSlite \cite{Agnese:2013jaa}, and SuperCDMS \cite{Agnese:2014aze}) 
As we can observe, there are plenty of points in the parameter space for the whole mass range that survive.
Remarkably, the predictions spread over the entire area of the parameter space which will be covered by direct detection experiments in the future. 
These theoretical predictions motivate future direct searches with low threshold experiments, such as the 
proposed future phase of SuperCDMS.

As shown in Ref.\,\cite{Cerdeno:2011qv}, there is in general a correlation between the WIMP annihilation cross-section in fermion-antifermion channels and the WIMP scattering cross-section off quarks that results from the crossing symmetry of the diagrams involved (and that is generic to any light DM candidate). 
This effect explains the points with large values of $\sigsi$ in the $b\bar b$ and $c\bar c$ channels. 
However, the correlation is broken if resonant effects are present in the annihilation cross-section, since the scattering cross-section does not present such resonances (and consequently can be much smaller).
As already observed in Fig.\,\ref{fig:mh1}, the points that survive the direct detection constraints for the $b\bar b$ and $c\bar c$ channels are precisely those in which resonant annihilation takes place (when $\snmassr\approx\hmassl/2$) and for this reason the predicted $\sigsi$ can decrease by several orders of magnitude.  This also explains the smallness of $\sigsi$ for the points in which annihilation in $gg$ dominates.

Similarly when annihilation into $\higgsl\higgsl$ or $\phiggsl\phiggsl$ channels dominates, the above mentioned correlation does not hold and $\sigsi$ can be very small as confirmed by Fig.\,\ref{fig:spin-independent}. 
The points with larger values of $\sigsi$ correspond to those in which the up component of the lightest (scalar or pseudoscalar) Higgs is large and the coupling to the $u$ quark increases.

Finally, black circles on the right-hand side of Fig.~\ref{fig:spin-independent} denote points for which the RH sneutrino annihilation cross-section exceeds the Fermi LAT bounds on dSph galaxies. 
Although this constraint is not very severe, we observe that it excludes some points which are not yet probed by direct detection. Its effect is more stringent for light WIMPs and many (but not all) of the solutions with $\snmassr\lesssim 5$~GeV can be excluded in this way. 
From the comparison of Fig.\,\ref{fig:spin-independent} and Fig.\,\ref{fig:sigmav} we can notice that indirect detection can probe points which would be very difficult (or impossible) to test with direct detection, and vice versa. This perfectly illustrates the need for complementary techniques in order to explore the whole DM parameter space.

\begin{figure}[t!]
	\begin{center}  
	\epsfig{file=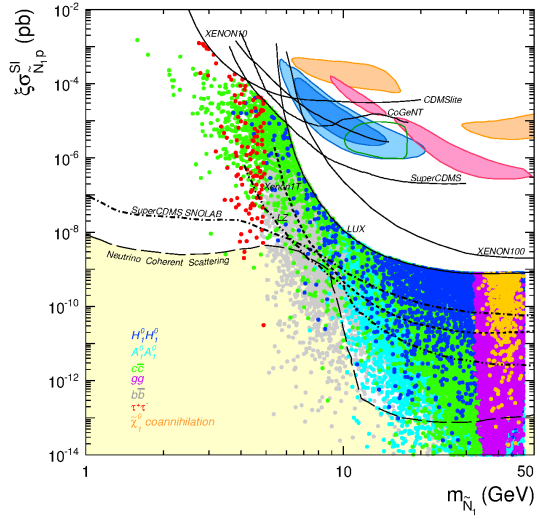,width=7.5cm}
	\end{center}
\vspace*{-0.5cm}
\caption{\small Theoretical predictions for $\sigsi$ as a function of the RH sneutrino mass. All the experimental constraints are included, together with the bounds from direct detection experiments and Fermi LAT data on dSphs. The colour code is as in Fig.\,\ref{fig:mh1}. Points corresponding to different annihilation channels are shown separately in Fig.\,\ref{fig:spin-independent-s}.
}
  \label{fig:sigsi}
\end{figure}

\begin{figure}[t!]
	\begin{center}  
	\hspace*{-0.5cm}\epsfig{file=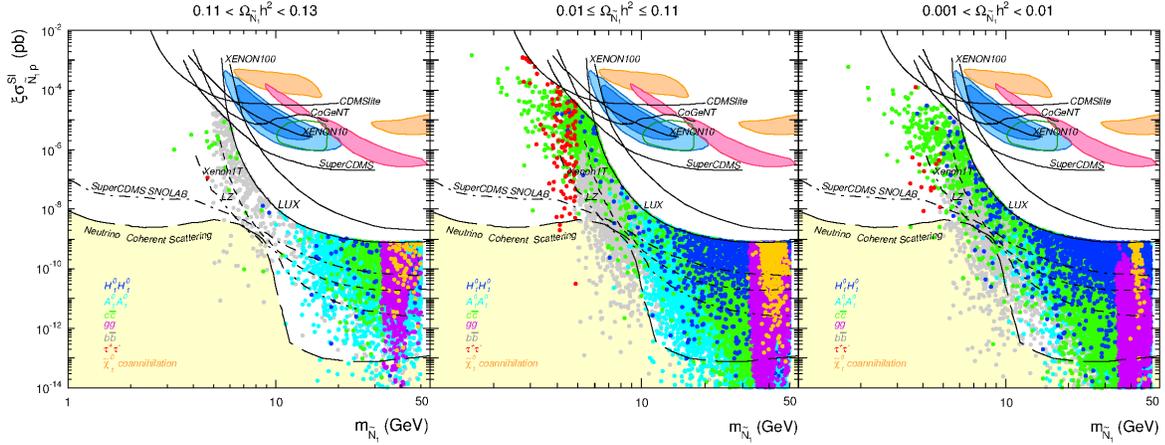,width=16cm}
	\end{center}
\vspace*{-0.5cm}
\caption{\small Theoretical predictions for $\sigsi$ as a function of the RH sneutrino mass. From left to right, the three panels show different ranges in the RH sneutrino relic abundance, the leftmost one being in agreement with Planck measurements.
All the experimental constraints are included, together with the bounds from direct detection experiments and Fermi LAT data on dSphs. The colour code is as in Fig.\,\ref{fig:mh1}. 
}
  \label{fig:sigsi_relic}
\end{figure}

For clearness, and as a comprehensive summary of our results, we represent in Fig.\,\ref{fig:sigsi} the theoretical predictions for $\sigsi$ after all the constraints have been applied, 
including those from direct and indirect searches.
The RH sneutrino in the NMSSM is therefore a very flexible candidate for light WIMP dark matter, that could account for future signals in this interesting mass region. Lastly, in Fig.\,\ref{fig:sigsi_relic} we have separated the results according to the relic abundance of RH sneutrinos. As already stressed in Section\,\ref{sec:indirect}, many of the examples with mass below 5~GeV have a small relic abundance. Nevertheless, they could be observable in future direct detection experiments.
We provide these results, including the dominant annihilation channels, as a data table available in \cite{database}.

\section{Conclusions}
\label{sec:conclusions}

We have investigated the viability of the RH sneutrino in the NMSSM as a candidate for light WIMP dark matter in the light of the recent experimental hints and constraints for this kind of candidates that arise from direct and indirect DM searches as well as from LHC data.

We have carried out a scan in the parameter space of the model, specifically looking for RH sneutrinos in the mass range $1- 50$~GeV that satisfy the upper limit on its relic abundance from Planck data. 
The most recent experimental constraints on low energy observables have been incorporated. These comprise the branching ratios of the rare processes $\bsg$, $\bmumu$,  and $B^+ \to \tau^+ \nu_\tau$, and the supersymmetric contribution to the muon anomalous magnetic moment, $\asusy$.
We have also taken into account the current lower bounds on the mass of SUSY particles. Regarding the Higgs sector, we have imposed the presence of a SM-like Higgs boson. The LHC measurements of the Higgs couplings lead to an upper bound on its invisible branching ratio, for which we consider ${\rm BR}(\hsm\rightarrow {\rm inv})<0.27$.
In order to reproduce the correct relic density while satisfying these bounds, the presence of a lighter singlet-like Higgs is needed. 
The extended structure of the NMSSM Higgs sector, together with the flexibility of the RH sneutrino parameters, make it possible to obtain viable RH sneutrino DM with a mass as light as 2~GeV.

We have computed the RH sneutrino annihilation cross section in the galactic halo, $\langle\sigma v\rangle_0$, and confronted it with the constraints that can be derived from Fermi LAT observations of the gamma ray flux from dwarf spheroidal galaxies. A small part of the parameter space is disfavoured by this bound, mostly corresponding to $\tau^+\tau^-$, $b\bar b$, and $c\bar c$ final states in RH sneutrino annihilation.  
Interestingly, we have also obtained specific examples where the RH sneutrino can have a mass $\snmassr=8-50 $~GeV and annihilate mostly into $b\bar b$, thereby accounting for the apparent low-energy excess in the gamma-ray emission at the GC identified in the Fermi LAT data.

Finally, the theoretical predictions for the RH sneutrino-proton elastic scattering cross section, $\sigsi$, have been computed and compared with current data form direct detection experiments. We have found regions of the parameter space for which $\sigsi$ can be compatible with some of the hints for light DM (e.g., CDMS-SI or CoGeNT).
After implementing the most recent upper bounds from CDMSlite, SuperCDMS and LUX, the predicted $\sigsi$ still spans various orders of magnitude for the mass range $\snmassr\approx2-50$~GeV, covering the whole parameter space and that can be explored by projected experiments. 
The flexibility of these predictions for low mass WIMPs constitute an excellent motivation for future low-threshold direct detection experiments.

\noindent{\bf \large Acknowledgements}

D.G.C. is supported by the Ram\'on y Cajal program of the Spanish MICINN. 
M.P. is supported by a MultiDark Scholarship.
We thank the support of the Consolider-Ingenio 2010 program under grant MULTIDARK CSD2009-00064, the Spanish MICINN under Grant No. FPA2012-34694, the Spanish MINECO ``Centro de excelencia Severo Ochoa Program" under Grant No. SEV-2012-0249, and the Community of Madrid under Grant No. HEPHACOS S2009/ESP-1473.

\appendix
\section{Additional plots}

For clarity, we include here some of the figures that have appeared in the main text, but where the points corresponding to different annihilation channels are shown separately.

\label{sec:}
\begin{figure}[t!]
	\begin{center}  
	\epsfig{file=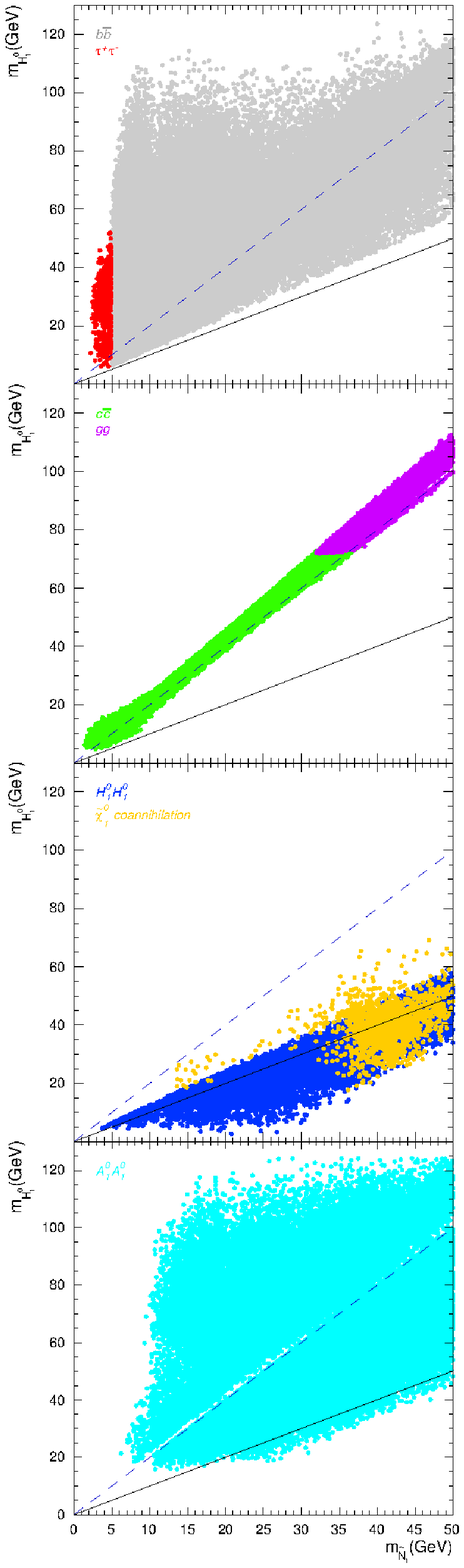,width=6.35cm}
	\epsfig{file=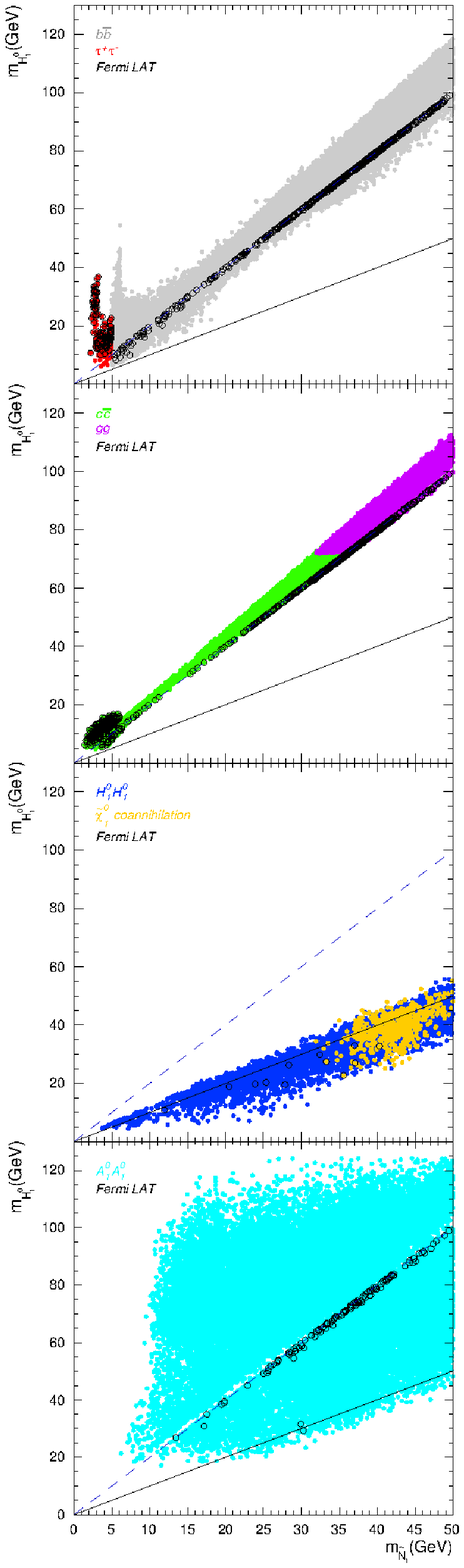,width=6.35cm}
	\end{center}
\caption{\small Same as Fig.\,\ref{fig:mh1} but the different annihilation channels are shown separately. 
}
  \label{fig:mh1-s}
\end{figure}

\begin{figure}[t!]
	\begin{center}  
	\epsfig{file=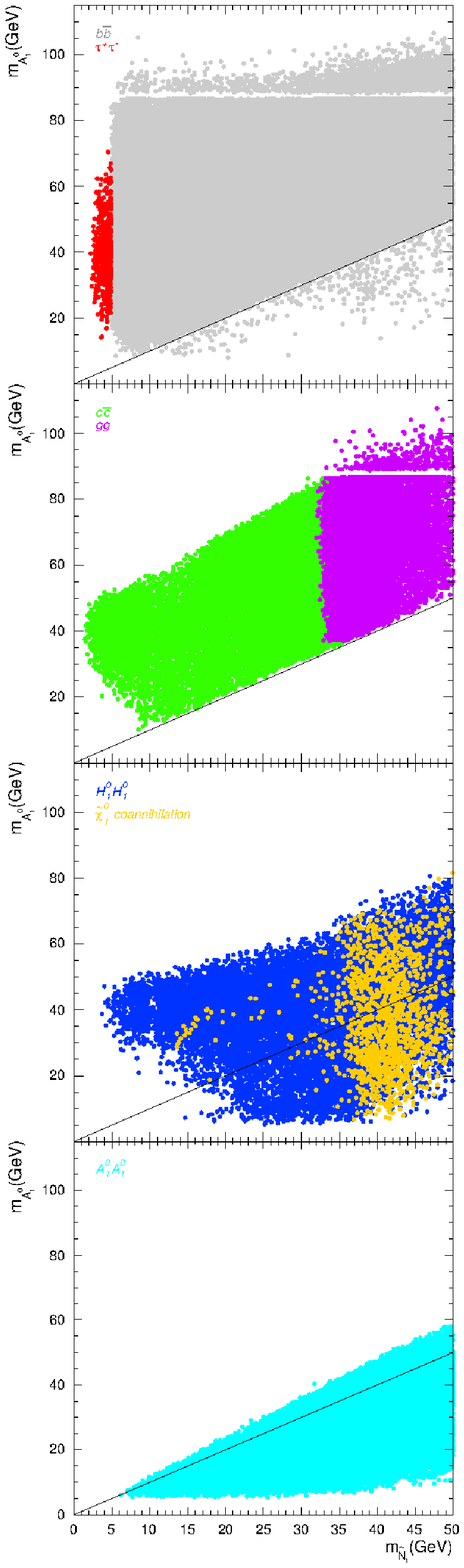,width=6.35cm}
	\epsfig{file=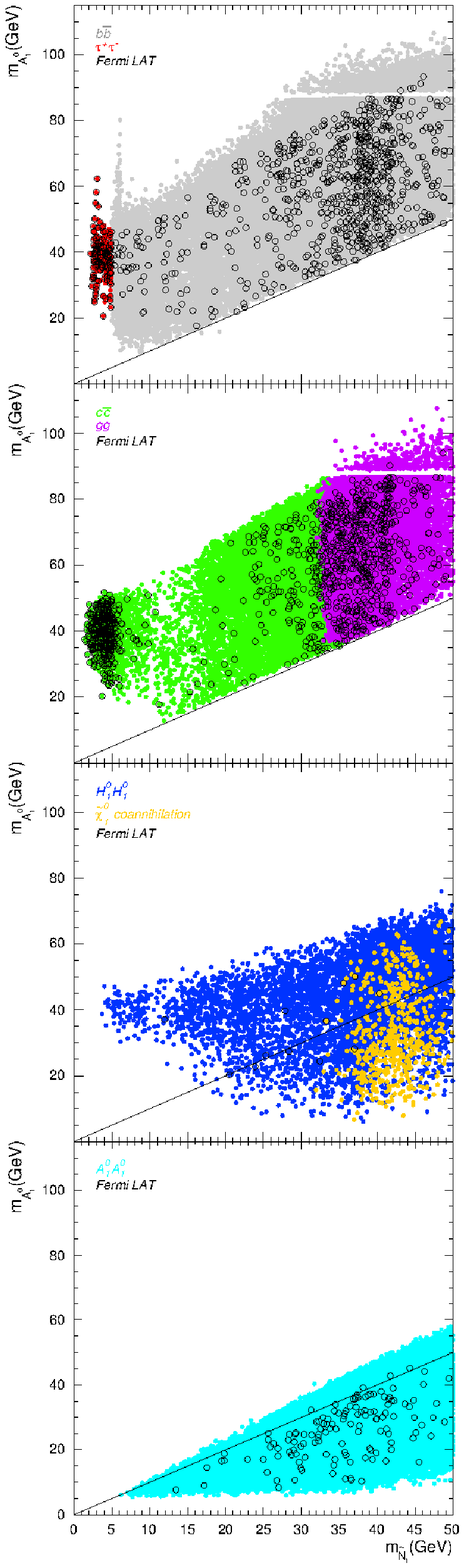,width=6.35cm}
	\end{center}
\caption{\small Same as Fig.\,\ref{fig:ma1} but the different annihilation channels are shown separately. 
 }
  \label{fig:ma1-s}
\end{figure}

\begin{figure}[t!]
	\begin{center}  
	\epsfig{file=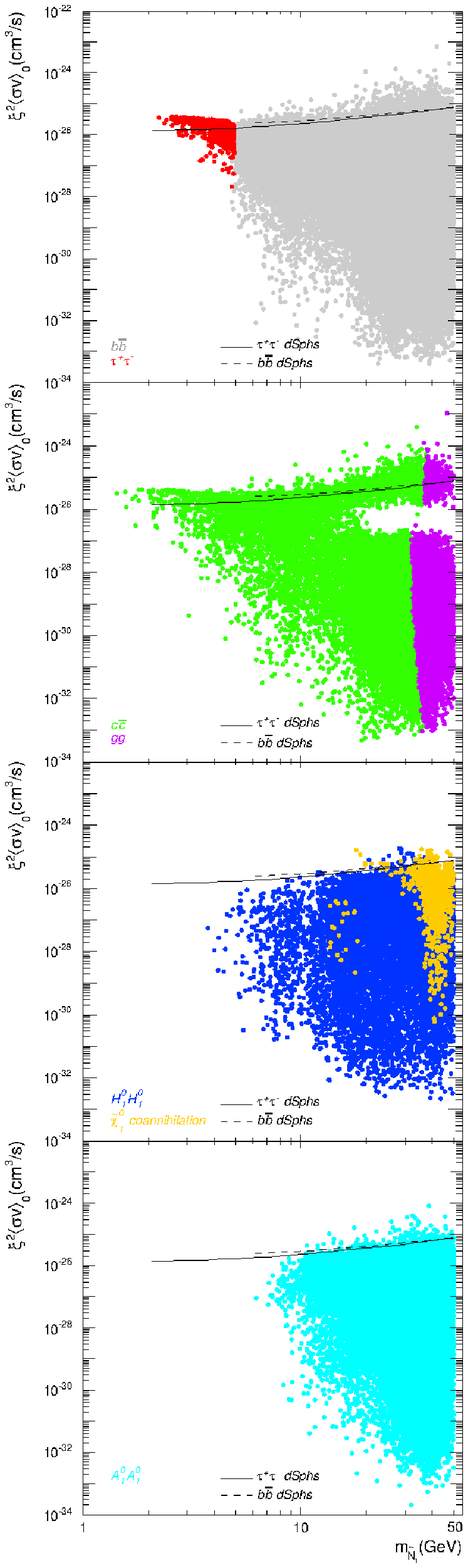,width=6.35cm}
	\epsfig{file=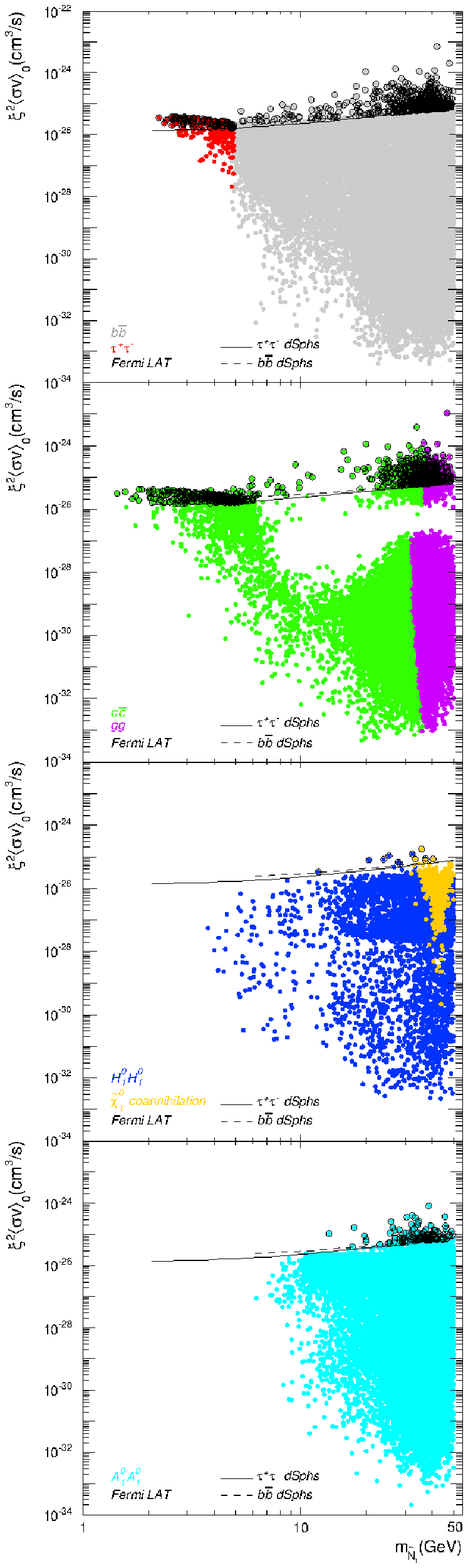,width=6.35cm}
	\end{center}
\caption{\small Same as Fig.\,\ref{fig:sigmav} but the different annihilation channels are shown separately. 
}
  \label{fig:sigmav-s}
\end{figure}

\begin{figure}[t!]
	\begin{center}  
	\epsfig{file=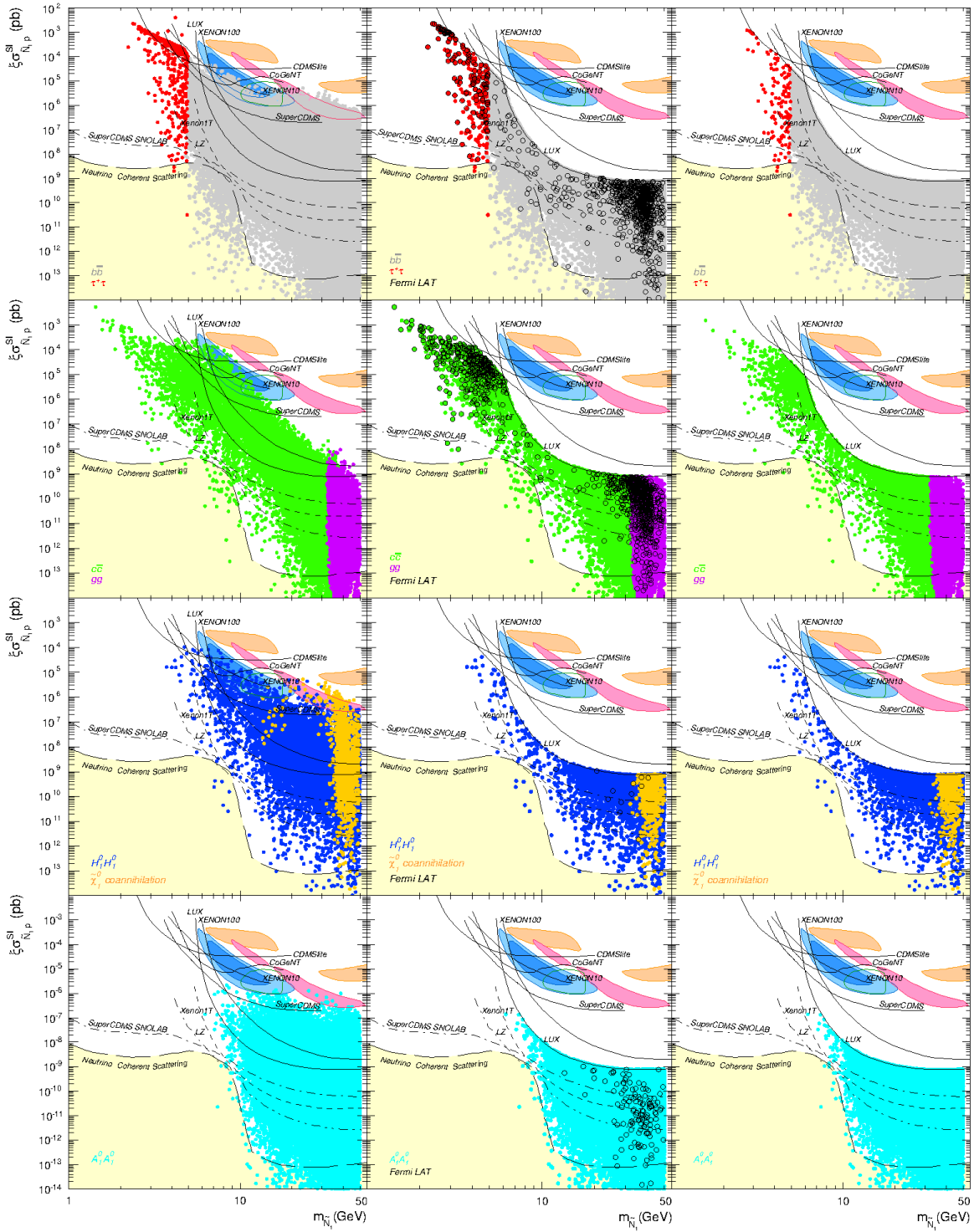,width=15.7cm}
	\end{center}
\caption{\small Same as Figs.\,\ref{fig:spin-independent} (left and middle panels) and \ref{fig:sigsi} (right panel) but the different annihilation channels are shown separately. 
}
  \label{fig:spin-independent-s}
\end{figure}

\end{document}